\documentstyle[preprint,cite,epsf,aps]{revtex}

\tightenlines
\newcommand{\be}{\begin{equation}}
\newcommand{\ee}{\end{equation}}
\newcommand{\bea}{\begin{eqnarray}}
\newcommand{\eea}{\end{eqnarray}}
\newcommand{\bml}{\begin{mathletters}}
\newcommand{\eml}{\end{mathletters}}
\newcommand{\pa}{\partial}

\newcommand{\e}{\epsilon}
\newcommand{\ve}{\varepsilon}

\begin{document}
\draft
\title{Noncommutative Quantum Field Theories\footnote{Lectures delivered at the XII Sumer School Jorge Andre Swieca. Section Particles and Fields, Campos de Jordan, SP, Brazil (2003).}}
\author{H. O. Girotti}
\address{Instituto de F\'{\i}sica,
Universidade Federal do Rio Grande do Sul \\ Caixa Postal 15051, 91501-970  - Porto Alegre, RS, Brazil\\ E-mail: hgirotti@if.ufrgs.br}

\maketitle

\begin{abstract}
We start by reviewing the formulation of noncommutative quantum mechanics as a constrained system. Then, we address to the problem of field theories defined on a noncommutative space-time manifold. The Moyal product is introduced and the appearance of the UV/IR mechanism is exemplified. The emphasis is on finding and analyzing noncommutative quantum field theories which are renormalizable and free of nonintegrable infrared singularities. In this last connection we give a detailed discussion of the quantization of the noncommutative Wess-Zumino model as well as of its low energy behavior. 

\end{abstract}
\newpage

\newpage

\begin{center}
{\bf Chapter I: NONCOMMUTATIVE QUANTUM MECHANICS}
\end{center}

\section{Introduction}
\label{sec:level1}

This set of lectures is mainly concerned with the problem of obtaining renormalizable quantum field theories defined on a noncommutative space-time manifold. However, as a first step into this problem, we would like to pin point the main features of quantum mechanics in a noncommutative space\footnote{For this chapter, and unlike in the remaining of these lectures, we adopt the cgs system of units}. Our line of development is based on the fact that a noncommutative geometry for the position variables arises, in some cases, from the canonical quantization of a dynamical system exhibiting second class constraints\cite{Deriglazov1}. 

To see how this come about we start by considering a nonsingular physical system whose configuration space is spanned by the coordinates $q_j, j = 1, 2, ...,N,$ and whose action ($S$) is 

\be
\label{I-1}
S[q]\,=\,\int dt \,L(q_j, \dot{q}_j)\,,
\ee

\noindent
where $L$ is the Lagrangian and $\dot{q}_j$ denotes the derivative of $q_j$ with respect to time. The dynamics, in the Lagrangian formulation, is controlled by the Lagrange equations of motion $\frac{\delta S[q]}{\delta q_j(t)} = 0$. 

For the quantization of the system under analysis, we must first switch into the Hamiltonian formulation of the classical dynamics. To this end, we first introduce $p_j$, the momentum canonically conjugate to $q_j$, as 

\be
\label{I-2}
p_j\,\equiv\,\frac{\pa L}{\pa \dot{q_j}}\,.
\ee

\noindent
The Hamiltonian ($H$) emerges from the Legendre transformation\footnote{We emphasize that the absence of constraints secures that (\ref{I-2}) is fully invertible.} 

\be
\label{I-3}
H[q , p]\,\equiv\,p_j\,\dot{q}_j\,-\,L(q_j, \dot{q}_j)\,.
\ee

Here, and everywhere else, repeated indices are to be summed. From (\ref{I-1}) and (\ref{I-3}) follows that the action can also be cast as a functional of coordinates and momenta, 

\be
\label{I-4}
S[q , p]\,=\,\int dt \left[ p_j\,\dot{q}_j\,-\,H(q_j, p_j)\right]\,.
\ee

\noindent
The dynamics in the Hamiltonian formulation is determined by solving the Hamilton equations of motion 

\bml
\label{I-401}
\bea
&&\frac{\delta S[q , p]}{\delta p_j(t)}\, = \,0\,\Longrightarrow \dot{q_j}\,=\,[q_j , H]_{PB}\,,
\label{mlett:I-a401}\\
&&\frac{\delta S[q , p]}{\delta q_j(t)}\, = \,0\,\Longrightarrow \dot{p_j}\,=\,[p_j , H]_{PB}\,,
\label{mlett:I-b401}
\eea
\eml

\noindent
where PB denotes Poisson brackets.

Canonical quantization consists in replacing $q_j \rightarrow Q_j$, $p_j \rightarrow P_j$, where $Q_j, j = 1, 2,...N$ and $P_j, j = 1, 2,...N$ are self-adjoint operators obeying the equal-time commutation algebra $[Q_j , Q_k] = 0, [P_j , P_k] = 0, [Q_j , P_k] = i\hbar \delta_{jk}$. This algebra is isomorphic to the PB algebra obeyed by the classical counterparts and is abstracted from it via the quantization rule

\be
\label{I-5}
i\hbar\,[,]_{PB}\biggr|_{
\begin{array}{c}
q \rightarrow Q \\ 
p \rightarrow P
\end{array}} 
\longrightarrow [,]\,.
\ee

\noindent
When ordering problems are absent, the composite operator $H(Q_j , P_j)$  is directly abstracted from its classical analog $H(q_j , p_j)$.   

As observed in Ref.\cite{Deriglazov1}, the description of the dynamics of a regular system is by no means unique. One may, for instance, enlarge the configuration space by adding the coordinates $v_j, j = 1, 2, ...N$ if, at the same time, one replaces the action in (\ref{I-1}) by the first order action

\be
\label{I-6}
S_1[q , v]\,=\,\int dt\,L_1(q_j , v_j)\,=\,\int dt\,\left[ v_j \dot{q}_j\,-\,H(q_j,v_j) \right]\,.
\ee

\noindent
The defining equations of the canonical conjugate momenta,

\bml
\label{I-7}
\bea
&& p_j\,\equiv\,\frac{\pa L_1}{\pa \dot{q_j}}\,=\,v_j\,, \label{mlett:aI-7}\\
&& \pi_j\,\equiv\,\frac{\pa L_1}{\pa \dot{v_j}}\,=\,0 \label{mlett:bI-7}\,,
\eea
\eml

\noindent
are not longer invertible but give rise to the primary constraints\footnote{ The sign of weak identity ($\approx$) is used in the sense of Dirac\cite{Dirac1}. As far as constrained systems are concerned, our terminology is strictly that put forward by Dirac\cite{Dirac1}.}  

\bml
\label{I-8}
\bea
&& G_j\,\equiv\,p_j \,-\,v_j\,\approx \,0\,,\label{mlett:aI-8}\\
&& T_j\,\equiv\,\pi_j\,\approx\,0\,,\label{mlett:bI-8}
\eea
\eml

\noindent
which verify the PB algebra

\bml
\label{I-9}
\bea
&& \left[ G_j , G_k \right]_{PB}\,=\,0 \,,\label{mlett:aI-9}\\
&& \left[ G_j , T_k \right]_{PB}\,=\,-\,\delta_{jk} \,,\label{mlett:bI-9}\\
&& \left[ T_j , T_k \right]_{PB}\,=\,0 \,.\label{mlett:cI-9}
\eea
\eml

\noindent
As for the canonical Hamiltonian it is found to read

\be
\label{I-10}
H_1\,=\,H(q_j , v_j)\,.
\ee

Since the primary constraints are already second-class, the Dirac algorithm\cite{Dirac1}\cite{Fradkin1}\cite{Sundermeyer1}\cite{Girotti1}\cite{Gitman1} does not yield secondary constraints. The Dirac brackets (DB) can be computed at once and one finds

\bml
\label{I-11}
\bea
&&\left[ q_j , q_k \right]_{DB}\,=\,0\,,\label{mlett:aI-11}\\
&&\left[ q_j , p_k \right]_{DB}\,=\,\delta_{jk}\,,\label{mlett:bI-11}\\
&&\left[ p_j , p_k \right]_{DB}\,=\,0\,.\label{mlett:cI-11}
\eea
\eml

\noindent
The DB's involving the remaining variables ($v_j, \pi_j$) can easily be obtained from those in Eq.(\ref{I-11}). Indeed, within the DB algebra the constraints hold as strong identities\cite{Dirac1} and one can therefore replace, wherever needed, $v_j$ by $p_j$ and $\pi_j$ by $0$. Thus, the sector $v_j, \pi_j, j = 1, 2, ...,N$ can be entirely eliminated and one is left with the so called physical phase space\cite{Fradkin1} ($\Gamma^{\ast}$), which is spanned by the variables $q_j , p_j, j = 1, 2, ...,N$. Furthermore, for these last mentioned set of phase space variables the DB's reduce to the corresponding PB's\cite{Fradkin1}, as confirmed by Eq.(\ref{I-11}). Then, the equations of motion deriving from

\bml
\label{I-12}
\bea
\dot{q}_j\,=\,\left[q_j , H_1\right]_{DB}\,,\label{mlett:aI-12}\\
\dot{p}_j\,=\,\left[q_j , H_1\right]_{DB}\,,\label{mlett:bI-12}
\eea
\eml

\noindent
are identical to those in Eq.(\ref{I-41}). Both descriptions are in fact equivalent.

\section{A noncommutative version of a regular system}
\label{sec:level2}

A noncommutative version of the system in Section I can be obtained modifying the action in (\ref{I-6}) by the addition of a Chern-Simons like term as follows\cite{Deriglazov1}

\be
\label{I-13}
S_N[q , v]\,=\,\int dt\,L_N(q_j , v_j)\,=\,\int dt\,\left[ v_j \dot{q}_j\,-\,H(q_j,v_j)\,+\,
\dot{v}_j \theta_{jk} v_k \right]\,,
\ee

\noindent
where $\|\theta\|$ is a nonsingular antisymmetric $N \times N$ matrix. One easily verifies that the primary constraints are now given by

\bml
\label{I-14}
\bea
&& G_j\,\equiv\,p_j \,-\,v_j\,\approx \,0\,,\label{mlett:aI-14}\\
&& T_j\,\equiv\,\pi_j\,-\,\theta_{jk} v_k\,\approx\,0\,,\label{mlett:bI-14}
\eea
\eml

\noindent
while the constraint algebra is found to be\footnote{Just for clarification we recall that in the cgs system of units the dimensions of the variables spanning the phase space are $d[q] = d[\pi] = cm$, while $d[p] = d[v] = gm\,cm\,sec^{-1}$. Furthermore, $d[\hbar] = gm\,cm^2\,sec^{-1}$. Thus, $d[\theta] = gm^{-1}\,sec$.} 

\bml
\label{I-15}
\bea
&& \left[ G_j , G_k \right]_{PB}\,=\,0 \,,\label{mlett:aI-15}\\
&& \left[ G_j , T_k \right]_{PB}\,=\,-\,\delta_{jk} \,,\label{mlett:bI-15}\\
&& \left[ T_j , T_k \right]_{PB}\,=\,-\,\theta_{jk} \,.\label{mlett:cI-15}
\eea
\eml

\noindent
In turns, this gives origin to the Dirac brackets

\bml
\label{I-16}
\bea
&&\left[ q_j , q_k \right]_{DB}\,=\,- 2\,\theta_{jk}\,,\label{mlett:aI-16}\\
&&\left[ q_j , p_k \right]_{DB}\,=\,\delta_{jk}\,,\label{mlett:bI-16}\\
&&\left[ p_j , p_k \right]_{DB}\,=\,0\,,\label{mlett:cI-16}
\eea
\eml

\noindent
which together with the canonical Hamiltonian

\be
\label{I-17}
H_N\,=\,H(q_j , v_j)\,,
\ee

\noindent
lead to the equation of motions

\bml
\label{I-18}
\bea
&&\dot{q}_j\,=\,\left[q_j , H_N\right]_{DB}\,=\,\frac{\pa H}{\pa p_j}\,-\,2\,\theta_{jk}\,\frac{\pa H}{\pa q_k}\,,\label{mlett:aI-18}\\
&&\dot{p}_j\,=\,\left[q_j , H_N\right]_{DB}\,=\,-\,\frac{\pa H}{\pa q_j}\,.\label{mlett:bI-18}
\eea
\eml

As before, we have used the constraints to eliminate from the game the sector $v_j, \pi_j, j = 1, 2, ...,N$. However, we can not refer to $q_j, p_j, j = 1, 2, ...,N$ as to the physical phase space coordinates because their DB's (see Eq.(\ref{I-16})) differ from the corresponding PB's. To find the physical phase space coordinates we follow Ref.\cite{Deriglazov1} and introduce

\bml
\label{I-19}
\bea
&&\tilde{q}_j\,\equiv\,q_j\,-\,\theta_{jk}\,p_k\,,\label{mlett:aI-19}\\
&&\tilde{p}_j\,\equiv\,p_j\,,\label{mlett:bI-19}
\eea
\eml 

\noindent
which fulfill, as desired,

\bml
\label{I-20}
\bea
&&\left[ \tilde{q}_j , \tilde{q}_k \right]_{DB}\,=\,\left[ \tilde{q}_j , \tilde{q}_k \right]_{PB}\,=\,0\,,\label{mlett:aI-20}\\
&&\left[ \tilde{q}_j , \tilde{p}_k \right]_{DB}\,=\,\left[ \tilde{q}_j , \tilde{p}_k \right]_{PB}\,=\,\delta_{jk}\,,\label{mlett:bI-20}\\
&&\left[ \tilde{p}_j , \tilde{p}_k \right]_{DB}\,=\,\left[ \tilde{p}_j , \tilde{p}_k \right]_{PB}\,=\,0\,.\label{mlett:cI-20}
\eea
\eml

Quantization now follows along standard lines: $\tilde{q}_j \rightarrow \tilde{Q}_j$, $\tilde{p}_j \rightarrow \tilde{P}_j$, where $\tilde{Q}_j, j = 1, 2,...N$ and $\tilde{P}_j, j = 1, 2,...N$ are self-adjoint operators obeying the equal-time commutation algebra $[\tilde{Q}_j , \tilde{Q}_k] = 0, [\tilde{P}_j , \tilde{P}_k] = 0, [\tilde{Q}_j , \tilde{P}_k] = i\hbar \delta_{jk}$. As for the Hamiltonian the classical-quantum correspondence yields

\be
\label{I-21}
H(q_j , p_j)\,\longrightarrow \,H(\tilde{Q}_j + \theta_{jk} \tilde{P}_k , \tilde{P}_j)\,.
\ee

\noindent
For instance, if the classical Hamiltonian reads 

\be
\label{I-22}
H(q_j , p_j)\,=\,\frac{p_j p_j}{2 m}\,+\,V(q_j)\,,
\ee

\noindent
the corresponding quantum mechanical Hamiltonian will be given by

\be
\label{I-23}
H(\tilde{Q}_j + \theta_{jk} \tilde{P}_k , \tilde{P}_j)\,=\,\frac{\tilde{P}_j \tilde{P}_j}{2 m}\,+\,V(\tilde{Q}_j + \theta_{jk} \tilde{P}_k)\,.
\ee

\noindent
In the position representation $\{|\tilde{q}_1,..., \tilde{q}_j, ...,\tilde{q}_N>\}$, where $\tilde{P}_j \rightarrow -i \hbar \pa/\pa \tilde{q}_j$, the development of the system in time is controlled by the wave equation

\be
\label{I-24}
-\,\frac{\hbar^2}{2 m}\,\nabla_{\tilde{q}}^2\,\Psi(\tilde{q} , t)\,+\,V\left(\tilde{q}_j - i \hbar \theta_{jk} \frac{\pa}{\pa \tilde{q}_k}\right)\,\Psi(\tilde{q} , t)\,=\,i \hbar \frac{\pa \,\Psi(\tilde{q} , t)}{\pa t}\,.
\ee

Now, the second term in the left hand side of Eq.(\ref{I-24}) can be written as 

\be
\label{I-25}
V(\tilde{q})\,\left[ exp \left(- i \hbar\,\overleftarrow{\frac{\pa}{\pa \tilde{q}}_j}\,\theta_{jk}\,\overrightarrow{\frac{\pa}{\pa \tilde{q}}_k}\right)\right]\,\Psi(\tilde{q} , t)\,.
\ee

\noindent
To see how this come about we first notice that\cite{Menzinescu1}

\bea
\label{I-26}
&& V(\tilde{q})\,\left[ exp \left(- i \hbar\,\overleftarrow{\frac{\pa}{\pa \tilde{q}}_j}\,\theta_{jk}\,\overrightarrow{\frac{\pa}{\pa \tilde{q}}_k}\right)\right]\,\Psi(\tilde{q} , t)\,\nonumber\\
&& = \sum_{n = 0}^{\infty}\,\frac{(- i \hbar)^n}{n!}\,\left[\frac{\pa}{\pa \tilde{q}}_{j_1}
...\frac{\pa}{\pa \tilde{q}}_{j_n}\,V(\tilde{q})\right]\,\theta_{j_1 k_1}...\theta_{j_n k_n}\,\left[\frac{\pa}{\pa \tilde{q}}_{k_1}...\frac{\pa}{\pa \tilde{q}}_{k_n}\,\Psi(\tilde{q} , t)\right]\nonumber\\
&& = \sum_{n = 0}^{\infty}\,\frac{1}{n!}\,\left[\frac{\pa}{\pa \tilde{q}}_{j_1}
...\frac{\pa}{\pa \tilde{q}}_{j_n}\,V(\tilde{q})\right]\,\theta_{j_1 k_1}...\theta_{j_n k_n}\,\left[\tilde{P}_{k_1}...\tilde{P}_{k_n}\Psi(\tilde{q} , t)\right]\,.
\eea

\noindent
Then, we explore the Fourier tranform of $V(\tilde{q})$ to write,

\be
\label{I-27}
\left[\frac{\pa}{\pa \tilde{q}}_{j_1}
...\frac{\pa}{\pa \tilde{q}}_{j_n}\,V(\tilde{q})\right]\,=\,\left(\frac{i}{\hbar}\right)^n\,
\int\,\frac{d^N \tilde{k}}{(2 \pi \hbar)^{\frac{N}{2}}}\,\tilde{k}_{j_1}...\tilde{k}_{j_n}\,
e^{\frac{i}{\hbar}\,\tilde{q}\cdot \tilde{k}}\,V(\tilde{k})\,.
\ee

\noindent
By going back with (\ref{I-27}) into (\ref{I-26}) one obtains

\bea
\label{I-28}
&&V(\tilde{q})\,\left[ exp \left(- i \hbar\,\overleftarrow{\frac{\pa}{\pa \tilde{q}}_j}\,\theta_{jk}\,\overrightarrow{\frac{\pa}{\pa \tilde{q}}_k}\right)\right]\,\Psi(\tilde{q} , t)
\nonumber\\
&&=\left\{\int\,\frac{d^N \tilde{k}}{(2 \pi \hbar)^{\frac{N}{2}}}\,e^{\frac{i}{\hbar}\,\tilde{k}_j\cdot [\tilde{q}_j + \theta_{jk}\,\tilde{P}_k]}\,V(\tilde{k})\right\}\,\,\Psi(\tilde{q} , t)
\,=\,V\left(\tilde{q}_j - i \hbar \theta_{jk} \frac{\pa}{\pa \tilde{q}_k}\right)\,\Psi(\tilde{q} , t)\,,
\eea

\noindent
as proposed

On the other hand, Eq.(\ref{I-25}) is the Moyal product\cite{Filk1} $V(\tilde{q})\,\ast \,\Psi(\tilde{q} , t)$, which is algebraically isomorphic to the product of composite operators $V(Q)\,\Psi(Q , t)$ defined on the manifold $[Q_j , Q_k] = -2 i \hbar\, \theta_{jk}\,\neq\,0$. Hence, we are effectively implementing quantum mechanics in a noncommutative manifold. Needless to say, these last set of commutation relations represents the translation into the quantum regime of (\ref{mlett:aI-16}). The fact that 

\be
\label{I-29}
V(\tilde{q})\,\ast \,\Psi(\tilde{q} , t) = V\left(\tilde{q}_j - i \hbar \theta_{jk} \frac{\pa}{\pa \tilde{q}_k}\right)\,\Psi(\tilde{q} , t)\,,
\ee

\noindent
was already recognized in\cite{Susskind1}. The differences between a general noncommutative quantum mechanical system and its commutative counterpart have been stressed in \cite{Gamboa1}.

\section{An example: The noncommutative two-dimensional harmonic oscillator}
\label{sec:level3}

Quantum mechanics in a noncommutative plane has been considered in Ref.\cite{Gamboa2}. We shall restrict here to study the noncommutative two-dimensional harmonic oscillator of mass $M$ and frequency $\omega$. Therefore, the dynamics of the system is determined, according to (\ref{I-23}), by the Hamiltonian operator ($H$)

\be
\label{I-30}
H\,=\,\frac{1}{2M}\,\left[ \tilde{P}_j \tilde{P}_j\,+\,M^2 \omega^2\,\left(\tilde{Q}_j + \theta \ve_{jk} \tilde{P}_k\right)\left(\tilde{Q}_j + \theta \ve_{jl} \tilde{P}_l\right)\right]
\,\equiv\,\left(1\,+\,M^2 \omega^2 \theta^2\right)\,H_{\theta}\,,
\ee

\noindent
where

\be
\label{I-31}
H_{\theta}\,=\,\frac{1}{2M}\,\left( \tilde{P}_j \tilde{P}_j\,+\,M^2 \omega^2_{\theta}\,\tilde{Q}_j \, \tilde{Q}_j\,+\,2\,\theta\,M^2\,\omega^2_{\theta}\,\ve_{jk}\,\tilde{Q}_j \tilde{P}_k \right)\,
\ee

\noindent
and

\be
\label{I-32}
\omega^2_{\theta}\,\equiv\,\frac{\omega^2}{\left(1\,+\,M^2 \omega^2\,\theta^2\right)}\,.
\ee

\noindent
Presently, the antisymmetric matrix $\|\theta\|$ has been chosen to be 

\be
\label{I-33}
\theta_{jk}\,=\,\theta\,\ve_{jk}\,,
\ee

\noindent
where $\ve_{jk}$ is the two-dimensional fully antisymmetric Levi-Civita tensor, ($\ve_{12} = +1$) verifying $ \ve_{jk} \ve_{kl} = - \delta_{jl}$, and $\theta$ is a constant. This election for $\|\theta\|$ preserves rotational invariance. 

We introduce next the creation ($a^{\dagger}_j$) and anhilation operators ($a_j$) obeying the commutator algebra

\bml
\label{I-34}
\bea
&&\left[ a_j\,,\,a_k \right]\,=\,0\,,\label{mlett:aI-34}\\
&&\left[ a^{\dagger}_j\,,\,a^{\dagger}_k \right]\,=\,0\,,\label{mlett:bI-34}\\
&&\left[ a_j\,,\,a^{\dagger}_k \right]\,=\,\delta_{jk}\,,\label{mlett:cI-34}
\eea
\eml

\noindent
in terms of which $\tilde{Q}_j$, $\tilde{P}_j$ can be written 

\bml
\label{I-35}
\bea
&&\tilde{Q}_j\,=\,\frac{1}{\sqrt{2}}\,\left(\frac{\hbar}{M \omega_{\theta}}\right)^{\frac{1}{2}} 
\left(a^{\dagger}_j\,+\,a_j\right)\,,\label{mlett:aI-35}\\
&&\tilde{P}_j\,=\,\frac{1}{\sqrt{2}}\,\left(\hbar M \omega_{\theta}\right)^{\frac{1}{2}} 
\left(a^{\dagger}_j\,-\,a_j\right)\,.\label{mlett:bI-35}
\eea
\eml

\noindent
After replacing (\ref{I-35}) into (\ref{I-31}) one obtains

\be
\label{I-36}
H_{\theta}\,=\,\hbar \omega_{\theta}\,\left( N \, +\,I\,+\,2 \theta M \omega_{\theta}\,J_2\right)\,,
\ee

\noindent
where

\bml
\label{I-37}
\bea
&&N\,\equiv\,a^{\dagger}_j\,a_j\,,\label{mlett:aI-37}\\
&&J_2\,\equiv\,\frac{1}{2 \hbar}\,L\,=\,\frac{1}{2 \hbar}\,\ve_{jk}\,\tilde{Q}_j \tilde{P}_k\,=\,-\,\frac{i}{2}\,a^{\dagger}_j \ve_{jk} a_k\,,\label{mlett:bI-37}
\eea
\eml

\noindent
$L$ is the angular momentum operator and $I$ is the identity operator.

In the sequel, it will prove useful to recognize that the system under analysis possesses, as well as its commutative couterpart, the $SU(2)$ symmetry whose generators ($J_1$, $J_2$, $J_3$) and Casimir operator ($J^2$) are

\bml
\label{I-38}
\bea
&&J_1\,=\,\frac{1}{2}\,\left(a^{\dagger}_2\,a_1\,+\,a^{\dagger}_1\,a_2\right)\,,
\label{mlett:aI-38}\\
&&J_2\,=\,-\,\frac{i}{2}\,a^{\dagger}_j \ve_{jk} a_k\,=\,\frac{i}{2}\,\left(a^{\dagger}_2\,a_1\,-\,a^{\dagger}_1\,a_2\right),\label{mlett:bI-38}\\
&&J_3\,=\,\frac{1}{2}\,\left(a^{\dagger}_1\,a_1\,-\,a^{\dagger}_2\,a_2\right)\,,\label{mlett:cI-38}\\
&&J^2\,=\,J_k J_k\,=\,\frac{N}{2}\,\left( \frac{N}{2} + 1\right)\,.\label{mlett:dI-38}
\eea
\eml

\noindent
It is straightforward to verify that (\ref{I-34}) implies $[J_k , J_l] = i \ve_{klm} J_m$ and $[J^2 , J_k] = 0$, as it must be\footnote{$\ve_{klm}$ is the three-dimensional fully antisymmetric Levi-Civita tensor with $\ve^{123} = 1$.}. Since $[N , J_k] = 0$, it follows from (\ref{I-36}) that the energy eigenvalue problem reads

\be
\label{I-39}
H_{\theta} |j , m>\,=\,\hbar \omega_{\theta}\,\left( n \, +\,1\,+\,2 \theta \,m\,M\, \omega_{\theta}\right)\,|j , m>\,,
\ee

\noindent
where $|j , m>$ are the common eigenvectors of $J^2$ and $J_2$. They are labeled by the eigenvalues of $J^2$ and $J_2$, named $j$ and $m$, respectively. As is well known, 

\bml
\label{I-40}
\bea
&&j = 0, 1/2, 1, 3/2, ...\,,\label{mlett:aI-40}\\
&& -j \leq m \leq j\,,\label{mlett:bI-40}
\eea
\eml 

\noindent
while

\be
\label{I-41}
n\,=\,2\,j\,.
\ee

\noindent
In the commutative case ($\theta = 0$) the degeneracy of the $n$th energy level is $2j + 1 = n+ 1$. The degeneracy is lifted by the noncommutativity.

The construction of the states $\{|j , m>\}$ by acting with creation operators on a certain vacuum state goes through the standard procedure. One starts by introducing new creation ($A^{\dagger}_{\pm}$) and anhilation ($A_{\pm}$) operators defined as

\bml
\label{I-42}
\bea
&& A_{\pm}\,\equiv\,\frac{1}{\sqrt{2}}\,\left(a_1\,\mp\,i a_2\right)\,,\label{mlett:aI-42}\\
&& A^{\dagger}_{\pm}\,\equiv\,\frac{1}{\sqrt{2}}\,\left(a^{\dagger}_1\,\pm\,i a^{\dagger}_2\right)\,,\label{mlett:bI-42}
\eea
\eml

\noindent
which fulfill the commutator algebra

\bml
\label{I-43}
\bea
&&\left[ A_{\alpha}\,,\,A_{\beta}\right]\,=0\,,\label{mlett:aI-43}\\
&&\left[ A^{\dagger}_{\alpha}\,,\,A^{\dagger}_{\beta}\right]\,=0\,,\label{mlett:bI-43}\\
&&\left[ A_{\alpha}\,,\,A^{\dagger}_{\beta}\right]\,=\delta_{\alpha \beta}\,,\label{mlett:cI-43}
\eea
\eml

\noindent
where $\alpha$ and $\beta$ are $+$ or $-$. Then, it turns out that\cite{Messiah} 

\be
\label{I-44}
|n_{+} , n_{-}>\,=\,\frac{1}{\sqrt{n_{+}!} \sqrt{n_{-}!}}\,\left(A_{+}^{\dagger}\right)^{n_{+}}\,
\left(A_{-}^{\dagger}\right)^{n_{-}}\,|0 , 0>\,,
\ee

\noindent
are, for $n_{\pm}$ semipositive definite integers and $A_{\pm}|0 , 0> = 0$, a complete and normalizable set of common eigenstates of the Hermitean operators

\bml
\label{I-45}
\bea
&&N_{+}\,\equiv\,A_{+}^{\dagger} A_{+}\,,\label{mlett:aI-45}\\
&&N_{-}\,\equiv\,A_{-}^{\dagger} A_{-}\,.\label{mlett:bI-45}
\eea
\eml

\noindent
Since by construction $[N_{+} , N_{-}] = 0$ and, furthermore,

\bml
\label{I-46}
\bea
&&N\,=\,N_{+}\,+\,N_{-}\,,\label{mlett:aI-46}\\
&&J_2\,=\,\frac{1}{2}\,\left(N_{+}\,-\,N_{-}\right)\,,\label{mlett:bI-46}
\eea
\eml

\noindent
one concludes that the common eigenstates of energy and angular momentum can also be denoted by $|n_{+} , n_{-}>$. The relationships among the labels follow from (\ref{I-41}) and (\ref{I-46}) 

\bml
\label{I-47}
\bea
&&2 j\,=\,n_{+}\,+\,n_{-}\,,\label{mlett:aI-47}\\
&&2 m\,=\,n_{+}\,-\,n_{-}\,.\label{mlett:bI-47}
\eea
\eml

\noindent
Of course, instead of (\ref{I-44}) we may write\cite{Gamboa2}

\be
\label{I-48}
|j , m>\,=\,\frac{1}{\sqrt{(j + m)!} \sqrt{(j - m)!}}\,\left(A_{+}^{\dagger}\right)^{(j + m)}\,
\left(A_{-}^{\dagger}\right)^{(j - m)}\,|0 , 0>\,.
\ee

It is easy to convince oneself that the effective Lagrangian ($L_{eff}$) giving origin to the Hamiltonian in Eq.(\ref{I-30}) reads

\be
\label{I-49}
L_{eff}\,=\,\frac{1}{2} M \dot{\tilde{q}_j} \dot{\tilde{q}_j}\,-\,M^2 \omega^2 \theta \tilde{q}_j \ve_{jk} \dot{\tilde{q}_k}\,-\,\frac{M \omega^2}{2}\,{\tilde{q}_j} {\tilde{q}_j}\,.
\ee

\noindent
The second term in the right hand side of Eq.(\ref{I-49}) describes the interaction of an electrically charged particle (charge $e$) with a constant magnetic field ($B$). The components of the corresponding vector potential ($\vec{A}$) can be read off from (\ref{I-49}),

\be
\label{I-50}
A_j\,=\,\frac{M^2 \omega^2 \theta c}{e}\,\ve_{jk}\,\tilde{q}_k\,,
\ee

\noindent
and, hence, the magnetic field turns out to be

\be
\label{I-51}
B\,=\,\vec{\nabla} \times \vec{A}\,=\,-2\,\frac{M^2 \omega^2 \theta c}{e}\,=\,constant\,,
\ee

\noindent
where $c$ denotes the speed of light in vacuum. Thus, the noncommutative two-dimensional harmonic oscillator maps into the Landau problem\cite{Landau1}.

The thermodynamic functions associated with the noncommutative two-dimensional harmonic oscillator are also of interest. Consider this system in thermodynamic equilibrium with a heat reservoir at temperature $T$. Let us first calculate the partition function $Z(\mu)$,

\be
\label{I-52}
Z_{\theta}(\mu)\,=\,Tr\,e^{- \mu H}\,,
\ee

\noindent
where $\mu = 1/kT$ and $k$ is the Boltzmann constant. From Eqs.(\ref{I-30}), (\ref{I-32}), (\ref{I-39}) and (\ref{I-47}) follows that the energy eigenvalue $E_{\theta}(n_{+},n_{-})$, labeled by $n_{+}$, $n_{-}$, can be cast as

\be
\label{I-53}
E_{\theta}(n_{+},n_{-})\,=\,\hbar \omega\,\sqrt{\left(1 + M^2 \omega^2 \theta^2\right)}\,\left[\left(1 + M \omega \theta\right) n_{+}\,+\,\left(1 - M \omega \theta\right) n_{-}\,+\,1\right]\,.
\ee

\noindent
Then, the trace in Eq.(\ref{I-52}) is readily find to be 

\be
\label{I-54}
Z_{\theta}(\mu)\,=\,e^{- y_{\theta}}\,\left(\sum_{n_{+}=0}^{\infty}\,e^{- y_{\theta} s_{\theta}^{+} n_{+} }\right)\,\left(\sum_{n_{-}=0}^{\infty}\,e^{- y_{\theta} s_{\theta}^{-} n_{-}} \right)\,,
\ee

\noindent
where 

\bml
\label{I-55}
\bea
&&y_{\theta}\,\equiv\,\mu \hbar \omega_{\theta}\,\left(1 + M^2 \omega^2 \theta^2\right)\,,\label{mlett:aI-55}\\
&&s_{\theta}^{\pm}\,\equiv\,1\,\pm\,M \omega_{\theta} \theta\,.\label{mlett:bI-55}
\eea
\eml

\noindent
Both sums, in this last equation, can be expressed in terms of known functions. For instance\cite{Gradshteyn1} 

\be
\label{I-56}
\sum_{n_{+}=0}^{\infty}\,e^{- y_{\theta} s_{\theta}^{+} n_{+} }\,=\,\frac{1}{2}\,e^{\frac{y_{\theta} s_{\theta}^{+}}{2}}\,\frac{1}{\sinh \left(\frac{y_{\theta} s_{\theta}^{+}}{2}\right)}\,.
\ee

\noindent
Hence\cite{Gamboa2},

\be
\label{I-57}
Z_{\theta}(\mu)\,=\,\frac{1}{4\,\sinh \left(\frac{y_{\theta} s_{\theta}^{+}}{2}\right)\,\sinh \left(\frac{y_{\theta} s_{\theta}^{-}}{2}\right)}\,.
\ee

One is now in position of computing several interesting quantities. However, we shall restrict, for the present time, to evaluate the mean energy ($<E_{\theta}>$) of the noncommutative quantized oscillator, 

\be
\label{I-58}
<E_{\theta}>\,=\,-\,\frac{\pa \left( \ln Z_{\theta}(\mu)\right)}{\pa \mu}\biggr|_{\mu = \frac{1}{k T}}\,.
\ee

\noindent
After some algebra one arrives at

\be
\label{I-59}
<E_{\theta}>\,=\,\frac{\hbar \omega_{\theta}}{2}\,\left[ \frac{1}{s_{\theta}^{-}}\,\coth \left(\frac{y_{\theta} s_{\theta}^{+}}{2}\right)\,+\, \frac{1}{s_{\theta}^{+}}\,\coth \left(\frac{y_{\theta} s_{\theta}^{-}}{2}\right)\right]\,,
\ee

\noindent
which in the commutative limit reproduces, as it must, Planck's formula, 

\be
\label{I-60}
<E_{\theta=0}>\,=\,\hbar \omega\,\coth \left(\frac{\hbar \omega}{2 k T}\right)\,.
\ee

To study another limiting situations it will prove convenient to cast $<E_{\theta}>$ in the following form

\bea
\label{I-61}
<E_{\theta}>\,&=&\,\hbar \omega \left\{\sqrt{1 + z^2}\right.\nonumber\\
&&\left.+\,\frac{ \sqrt{1 + z^2}\,+\,z}{\exp\left[\frac{\hbar \omega}{k T} \left(\sqrt{1 + z^2}\,+\,z\right)\right] - 1}\,+\,\frac{ \sqrt{1 + z^2}\,-\,z}{\exp\left[\frac{\hbar \omega}{k T} \left(\sqrt{1 + z^2}\,-\,z\right)\right] - 1}\right\}\,,
\eea

\noindent
where $z$ is the dimensionless variable

\be
\label{I-62}
z\,\equiv\,M \omega \theta\,.
\ee

\noindent
1) Assume $\omega$ and $z$ fixed, while $T$ varies. Then,

\be
\label{I-63}
\lim_{\hbar \omega \ll k T} <E_{\theta}> \longrightarrow 2\,k\,T\,,
\ee

\noindent
implying that the noncommutativity does not alter the high temperature limit. Alternatively, at the other end of the scale temperature one finds that

\be
\label{I-64}
\lim_{\hbar \omega \gg k T} <E_{\theta}> \longrightarrow \hbar \omega \,\sqrt{1 + z^2}\,,
\ee

\noindent
which, as expected, coincides with the energy eigenvalue in Eq.(\ref{I-53}) for $n_{+} = n_{-} = 0$.

\noindent
2) Assume $\omega$, $M$, and $T$ fixed, while $\theta$ varies. In this case $z$ is just proportional to $\theta$, the proportionality constant ($M \omega$) being positive. Let us now investigate the limit of infinite noncommutativity specified by $\theta \longrightarrow +\infty \Longrightarrow z \longrightarrow +\infty$. At this limit, the behavior of those terms in (\ref{I-61}) containing an exponential factor in the denominator is quite different. In fact,

\bml
\label{I-65}
\bea
&&\lim_{z \rightarrow +\infty}\,\frac{ \sqrt{1 + z^2}\,+\,z}{\exp\left[\frac{\hbar \omega}{k T} \left(\sqrt{1 + z^2}\,+\,z\right)\right] - 1}\,\longrightarrow\,0\,,\label{mlett:aI-65}\\
&&\lim_{z \rightarrow +\infty}\,\frac{ \sqrt{1 + z^2}\,-\,z}{\exp\left[\frac{\hbar \omega}{k T} \left(\sqrt{1 + z^2}\,-\,z\right)\right] - 1}\,\longrightarrow\,\frac{k T}{\hbar \omega}\,.\label{mlett:bI-65}
\eea
\eml

\noindent
As can easily be seen, the situation is exactly reversed if $\theta \longrightarrow -\infty \Longrightarrow z \longrightarrow -\infty$. For both of these cases Eq.(\ref{I-61}) collapses into 

\be
\label{I-66}
\lim_{|\theta|\rightarrow +\infty} <E_{\theta}>\longrightarrow \hbar \omega \left( M \omega |\theta|\,+\,\frac{kT}{\hbar \omega}\right)\,\longrightarrow \,\hbar \omega^2 M |\theta|\,.
\ee 

\noindent
So, things look similar to the high temperature limit in the commutative case with a temperature given by 

\be
\label{I-67}
\frac{\hbar \omega^2 M}{k}\,|\theta|\,.
\ee

\newpage

\begin{center}
{\bf Chapter II: THE MOYAL PRODUCT. COMPUTATION OF VERTICES. THE UV/IR MECHANISM}
\end{center}

\section{Introduction}
\label{sec:level4}

In this Chapter we start by summarizing the main mathematical tools and results needed for studying field theories defined on a noncommutative space-time manifold\footnote{Hereafter we shall employ the natural system of units ($\hbar = 1$, $c = 1$).}. Then, we work out explicitly the problem of finding the Feynman rules for certain noncommutative field theories. At the end, we study the ultraviolet-infrared (UV/IR) mixing mechanism, which is the distinctive mark of noncommutative field theories. 

The first paper on quantum field theories formulated in a noncommutative space-time manifold was published in 1947\cite{Snyder1}, although the idea that a noncommtative space-time manifold might provide a solution for the problem of UV divergences seems to have been suggested long before\cite{Jackiw1}. The subject was abandoned, due the the success of renormalization theory, and its revival is rather recent and related to string theory. Indeed, the noncommutative Yang-Mills theory arises as a limit of string theory\cite{Connes1} and was extracted by Seiberg and Witten\cite{Seiberg1} starting from the open string in the presence of a magnetic ($B$) field. 

It has been stressed\cite{Deriglazov2} that this last mentioned system can be quantized by following stricly the Dirac quantization procedure\cite{Dirac1}, much as we did in Chapter I in connection with nonrelativistic systems. We shall, nevertheles, omit the details here and refer the reader to the original paper\cite{Deriglazov2}. More details on this and related subjects can be found in the already existing review articles\cite{Rivelles1,Douglas1,Szabo1,Gomes1}. 

\section{The Moyal product}
\label{sec:levelV}

Our starting point will then be the introduction of a $d$-dimensional noncommutative space-time. This is effectively done by declaring that time and position are not longer c-numbers but self-adjoint operators ($q^{\mu}, \mu = 0, 1, ...,d-1$) defined in a Hilbert space and obeying the commutation algebra\cite{Filk1}

\be
\label{II-1}
\left[ q^{\mu}\,,\,q^{\nu}\right]\,=\,i\,\Theta^{\mu \nu}\,,
\ee

\noindent
where the $\Theta^{\mu \nu}$'s are the elements of a real numerical $d \times d$ antisymetric matrix ($\| \Theta \|$) which, obviously, commutes with the $q$'s. One introduces next the operator $T(k)$

\be
\label{II-2}
T(k)\,\equiv\,e^{i\,k_{\mu}\,q^{\mu}}\,,
\ee

\noindent
where the $k$'s are c-numbers. The self-adjointness of the $q$'s implies that

\be
\label{II-3}
T^{\dagger}(k)\,=\,T(- k)\,,
\ee

\noindent
while $\left[q^{\mu}\,,\,\Theta^{\nu \rho}\right]=0$ leads to\footnote{Recall the Baker-Campbell-Hausdorf formula: $e^A e^B\,=\,e^{A+B}\, e^{\frac{1}{2}[A , B]}$ for $[A , B]$ a c-number.}

\be
\label{II-4}
T(k)\,T(k')\,=\,T(k + k')\,e^{- \frac{i}{2}\,k_{\mu} k'_{\nu}\,\Theta^{\mu \nu}}\,.
\ee

\noindent
Furthermore,

\be
\label{II-5}
tr T(k)\,=\,\left( 2 \pi \right)^d\,\prod_{\mu = 0}^{d - 1}\,\delta \left(k_{\mu}\right)\,,
\ee

\noindent
where the trace is taken with respect to a basis in representation space\cite{Filk1}.

We now follow Weyl\cite{Weyl1} and associate to a classical field $\phi(x)$ the operator $\Phi$ according to the rule

\be
\label{II-6}
\Phi\,=\,\frac{1}{(2 \pi)^d}\,\int \,dx\,dk\,T(k)\,e^{i k_{\mu} x^{\mu}}\,\phi(x)\,=\,
\frac{1}{(2 \pi)^d}\,\int \,dk\,T(k)\,\tilde{\phi}(k)\,,
\ee

\noindent
where $dx \equiv d^dx$, $dk \equiv d^dk$, and $\tilde{\phi}(k)$,

\be
\label{II-7}
\tilde{\phi}(k)\,\equiv\,\int\,dx\,e^{i k_{\mu} x^{\mu}}\,\phi(x)\,,
\ee

\noindent
is the Fourier transform of $\phi(x)$. The inverse mapping, $\Phi \longrightarrow \phi(x)$, is readily obtained by taking advantage of (\ref{II-5}). It reads

\be
\label{II-8}
\phi(x)\,=\,\int\,\frac{dk}{(2 \pi)^d}\,e^{ - i k_{\mu} x^{\mu}}\,tr\left[\Phi T^{\dagger}(k)\right]\,.
\ee

In the sequel, the Moyal product is introduced as

\be
\label{II-9}
\phi_1(x) \ast \phi_2(x)\,=\,\int\,\frac{dk}{(2 \pi)^d}\,e^{ - i k_{\mu} x^{\mu}}\,tr\left[\Phi_1 \Phi_2 T^{\dagger}(k)\right]\,,
\ee

\noindent
and, on general grounds,

\be
\label{II-10}
\phi_1(x) \ast \phi_2(x)\ast ... \ast \phi_n(x)\,=\,\int\,\frac{dk}{(2 \pi)^d}\,e^{ - i k_{\mu} x^{\mu}}\,tr\left[\Phi_1 \Phi_2 ... \Phi_n T^{\dagger}(k)\right]\,.
\ee

\noindent
After noticing that $T^{\dagger}(0) = I$ (see (\ref{II-2})) one concludes that 

\be
\label{II-11}
\int\,dx\,\phi_1(x) \ast \phi_2(x)\ast ... \ast \phi_n(x)\,=\,tr\left[\Phi_1 \Phi_2 ... \Phi_n \right]\,.
\ee

\noindent
Namely, the integral of the Moyal product turns out to be invariant under cyclical permutations of the fields. 

We shall be needing an alternative form of the Moyal product which exhibits, among other things, its highly nonlocal nature. The use of $(\ref{II-6})$ together with the defining properties of the operator $T(k)$ enables one to find, after some algebra,

\bea
\label{II-12}
&&tr\left[\Phi_1 \Phi_2 T^{\dagger}(k) \right]\,=\,\int\,\frac{dk_1}{(2 \pi)^d}\,\int \frac{dk_2}{(2 \pi)^d}\,\tilde{\phi(k_1)}\,\tilde{\phi(k_2)}\,tr\left[T(k_1) T(k_2)  T^{\dagger}(k) \right]\nonumber\\
&&=\,\int\,\frac{dk_1}{(2 \pi)^d}\,\int \frac{dk_2}{(2 \pi)^d}\,\tilde{\phi(k_1)}\,\tilde{\phi(k_2)}\,e^{- \frac{i}{2} k^{\mu}_1 k^{\nu}_2 \Theta_{\mu \nu}}\,(2 \pi)^d\,\delta(k_1 + k_2 - k)\,,
\eea

\noindent
where, unless otherwise specified, $\delta(k) = \prod_{\mu = 0}^{d}\,\delta \left(k_{\mu}\right)$. We use next the Fourier antitransform to replace, in the last term of the equality (\ref{II-12}), $\tilde{\phi(k)}$ in terms of $\phi(x)$. Once this has been done, we go with the resulting expression into (\ref{II-9}). After some algebraic rearrangements one arrives at the desired expression

\bea
\label{II-13}
&&\phi_1(x) \ast \phi_2(x)\,=\,\phi_1(x)\,exp \left(\frac{i}{2} \,\overleftarrow{\frac{\pa}{\pa x ^{\mu}}}\,\Theta^{\mu \nu}\,\overrightarrow{\frac{\pa}{\pa x^{\nu}}}\right)\,\phi_2(x)\nonumber\\
&&=\,\sum_{n = 0}^{\infty}\,\left(\frac{i}{2}\right)^n\,\frac{1}{n!}\,\left[\pa_{\mu_1} \pa_{\mu_2}...\pa_{\mu_n}\,\phi_1(x)\right]\,\Theta^{\mu_1 \nu_1} \Theta^{\mu_2 \nu_2} ... \Theta^{\mu_n \nu_n}\,\left[\pa_{\nu_1} \pa_{\nu_2}...\pa_{\nu_n}\,\phi_2(x)\right]\,,
\eea

\noindent
which explicitly displays the nonlocality of the Moyal product. Also, as a by product, one finds that

\be
\label{II-14}
\int\,dx\,\phi_1(x) \ast \phi_2(x)\,=\,\int\,dx\,\phi_1(x) \, \phi_2(x)\,,
\ee

\noindent
where we have assumed that all surface terms vanish. In words, under the integral sign the Moyal product of two fields reduces to the ordinary product.

Consider now the problem of quantizing  a noncommutative field theory within the perturbative approach. The first step into this direction consists in determining the corresponding Feynman rules. We shall only be dealing with field theories whose action, in the commutative counterpart, is composed of kinetic terms that are quadratic in the fields, plus interaction terms that are polinomials in the fields of degree higher than two. In the noncommutative situation all ordinary field products are replaced by the corresponding Moyal products. However, according to (\ref{II-14}), this replacement does not affect the kinetic terms which, in turns, implies that the propagators remain as in the commutative case. Only the vertices are modified by the noncommutativity. Our next task will, then, be to complete the determination of the Feynman rules by explicitly finding, in momentum space, the expressions for the vertices of some theories of interest.      

\section{Computation of vertices}
\label{sec:levelVI}

To this end we must start by evaluating the right hand side of (\ref{II-11}). According to (\ref{II-6}) we have that

\bea
\label{II-15}
tr\left[\Phi_1 \Phi_2 ... \Phi_n \right]\,&=&\,\int\,\left(\prod_{j = 1}^{n} dx_j\right)\,\left\{ \int\,\left(\prod_{j = 1}^{n} \frac{dk_j}{(2 \pi)^d}\right)\,e^{i k_{\mu}^j x^{\mu}_j}\,tr\left[T(k_1) T(k_2)...T(k_n)\right]\right\}\nonumber\\
&&\times\phi(x_1) \phi(x_2) ... \phi(x_n)\,.
\eea

\noindent
For the evaluation of the trace, in the right hand side of (\ref{II-15}), we merely need to explore the algebra obeyed by the $T$'s. One obtains

\be
\label{II-16}
tr\left[T(k_1) T(k_2)...T(k_n)\right]\,=\,\exp\left(- \frac{i}{2} \sum_{i<j}^n\,k_{\mu}^i k_{\nu}^j\,\Theta^{\mu \nu}\right)\,(2 \pi)^d\,\delta(k_1 + k_2 + ... + k_n)\,.
\ee

\noindent
Hence, from (\ref{II-16}), (\ref{II-15}), and (\ref{II-11}) one arrives at

\be
\label{II-17}
\int\,dx\,\phi_1(x) \ast \phi_2(x)\ast ... \ast \phi_n(x)\,=\,\int\,\left(\prod_{j = 1}^{n} dx_j\right)\,V(x_1, x_2, ..., x_n)\, \phi(x_1) \phi(x_2) ... \phi(x_n)\,,
\ee

\noindent
where we have introduced the definition

\bea
\label{II-18}
V(x_1, x_2, ..., x_n)\,&\equiv&\, \int\,\left[\prod_{j = 1}^{n} \frac{dk_j}{(2 \pi)^d}\right]\,e^{i k_{\mu}^j x^{\mu}_j}\,\exp\left(- \frac{i}{2} \sum_{i<j=1}^n\,k_{\mu}^i k_{\nu}^j\,\Theta^{\mu \nu}\right)\nonumber\\
&\times&\,(2 \pi)^d\,\delta(k_1 + k_2 + ... + k_n)\,.
\eea

\noindent
We may, if we wish, express the right hand side of (\ref{II-17}) in terms of Fourier transforms. One easily finds that

\be
\label{II-19}
\int\,dx\,\phi_1(x) \ast \phi_2(x) ... \ast \phi_n(x)\,=\,\int\,\left[\prod_{j = 1}^{n} \frac{dk_j}{(2 \pi)^d}\right]\,\tilde{V}(k_1, k_2, ..., k_n)\,\tilde{\phi}(k_1) \tilde{\phi}(k_2) ... \tilde{\phi}(k_n)\,,
\ee

\noindent
where

\be
\label{II-20}
\tilde{V}(k_1, k_2, ..., k_n)\,\equiv\,\exp\left(- \frac{i}{2} \sum_{i<j=1}^n\,k_{\mu}^i k_{\nu}^j\,\Theta^{\mu \nu}\right)\,(2 \pi)^d\,\delta(k_1 + k_2 + ... + k_n)\,.
\ee

\noindent
Observe that in the commutative case ($\Theta^{\mu \nu} = 0$) the scalar vertex reduces, as it must, to

\be
\label{II-21}
\tilde{V}(k_1, k_2, ..., k_n)\,=\,(2 \pi)^d\,\delta(k_1 + k_2 + ... + k_n)\,.
\ee

We analyze next some specific examples.

\noindent
1) The interaction Lagrangian is 

\be
\label{II-22}
{\cal L}_I\,=\,g\,\phi_1(x) \ast \phi_2(x)\ast \phi_3(x)\,.
\ee

\noindent
Here, $\phi_1$, $\phi_2$, and $\phi_3$ are three different scalar fields whereas and $g$ is a coupling constant. To find the vertex we shall start by looking for the lowest order perturbative contribution to the three-point connected Green function $<0|T(\phi_1(x_1)  \phi_2(x_2) \phi_3(x_3)\,\hat{S})|0>$, where $\hat{S}$ is the scattering operator and $T$ designates the chronological ordering operator. According to the rules of quantum field theory such contribution is given by $<0|T(\phi_1(x_1)  \phi_2(x_2) \phi_3(x_3)\,\hat{S}_1)|0>$, where

\bea
\label{II-23}
\hat{S}_1\,&=&\,i\,\int d^dx\,{\cal L}_I\,=\,i\,g\,\int d^dx\,\phi_1(x) \ast \phi_2(x)\ast \phi_3(x)\nonumber\\
&=&\,i\,g\,\int\, d^dx_1\,d^dx_2\,d^dx_3\,V(x_1, x_2, x_3)\,\phi_1(x_1)  \phi_2(x_2) \phi_3(x_3)\,.
\eea

\noindent
Clearly, we have used (\ref{II-17}) for arriving at the last term in the equality (\ref{II-23}). Having reached this point we invoke Wick's theorem to find

\bea
\label{II-24}
&&\int\,\left(\prod_{j = 1}^{3} dx_j\right)\,e^{i p^1_{\mu} x_1^{\mu} + i p^2_{\mu} x_2^{\mu} + i p^3_{\mu} x_3^{\mu}}\,<0|T(\phi_1(x_1)  \phi_2(x_2) \phi_3(x_3)\,\hat{S}_1)|0>\nonumber\\
&=&i \,g\,\tilde{G}_1(p_1)\,\tilde{G}_2(p_2)\,\tilde{G}_3(p_3)\,\int \,\left(\prod_{j = 1}^{3} dx_j\right)\,V(x_1, x_2, x_3)\,e^{i p^1_{\mu} x_1^{\mu} + i p^2_{\mu} x_2^{\mu} + i p^3_{\mu} x_3^{\mu}}\,.
\eea

\noindent
Notice that in the right hand side of this last equation we already have the Fourier transform of the connected three-point function we were looking for\footnote{All external momenta will always be assumed leaving the vertex}. Also, we have introduced the Fourier transforms ($\tilde{G}_j(p_j), j = 1, 2, 3$) of the two-point functions $<0|T(\phi_j(x_j) \phi_j(y_j))|0>$ which are assumed to be nonsingular. The expression for the vertex arises after amputating these two-point functions and, therefore, reads

\be
\label{II-25}
i\,g\,\int \,\left(\prod_{j = 1}^{3} dx_j\right)\,V(x_1, x_2, x_3)\,e^{i p^1_{\mu} x_1^{\mu} + i p^2_{\mu} x_2^{\mu} + i p^3_{\mu} x_3^{\mu}}\,,
\ee

\noindent
which after taking into account (\ref{II-18}) goes into

\be
\label{II-26}
i\,g\,\exp\left(- \frac{i}{2} \sum_{i<j=1}^3\,p_{\mu}^i p_{\nu}^j\,\Theta^{\mu \nu}\right)
\,(2 \pi)^d\,\delta(p_1 + p_2 + p_3)\,.
\ee

\noindent
The presence of the delta function, securing conservation of linear four-momentum at the vertex, together with the antisymmetric character of $\|\Theta\|$ allows to rewrite (\ref{II-26}) in the following final form

\be
\label{II-27}
i\,g\,(2 \pi)^d\,\delta(p_1 + p_2 + p_3)\,e^{- i \left(p_1 \wedge p_2\right)}\,,
\ee

\noindent
where

\be
\label{II-28}
p_1 \wedge p_2\,=\,\frac{1}{2}\,p_{\mu}^1\,p_{\nu}^2\,\Theta^{\mu \nu}\,.
\ee

\noindent
2) The interaction Lagrangian is 

\be
\label{II-29}
{\cal L}_I\,=\,\frac{g}{2}\,\phi_1(x) \ast \phi_1(x)\ast \phi_2(x)\,.
\ee

\noindent
This time Wick's theorem lead us to

\bea
\label{II-30}
&&\int\,\left(\prod_{j = 1}^{3} dx_j\right)\,e^{i p^1_{\mu} x_1^{\mu} + i p^2_{\mu} x_2^{\mu} + i p^3_{\mu} x_3^{\mu}}\,<0|T(\phi_1(x_1)  \phi_1(x_2) \phi_2(x_3)\,\hat{S}_1)|0>\nonumber\\
&=&\,i \,\frac{g}{2}\,\tilde{G}_1(p_1)\,\tilde{G}_1(p_2)\,\tilde{G}_2(p_3)
\nonumber\\
&\times&\,\int \,\left(\prod_{j = 1}^{3} dx_j\right)\,V(x_1, x_2, x_3)\,\left(e^{i p^1_{\mu} x_1^{\mu} + i p^2_{\mu} x_2^{\mu} + i p^3_{\mu} x_3^{\mu}}\,+\,e^{i p^1_{\mu} x_2^{\mu} + i p^2_{\mu} x_1^{\mu} + i p^3_{\mu} x_3^{\mu}}\right).
\eea

\noindent
After the truncation of the two-point functions the following expression for the vertex emerges

\be
\label{II-31}
i \,\frac{g}{2}\,\int \,\left(\prod_{j = 1}^{3} dx_j\right)\,V(x_1, x_2, x_3)\,\left(e^{i p^1_{\mu} x_1^{\mu} + i p^2_{\mu} x_2^{\mu} + i p^3_{\mu} x_3^{\mu}}\,+\,e^{i p^1_{\mu} x_2^{\mu} + i p^2_{\mu} x_1^{\mu} + i p^3_{\mu} x_3^{\mu}}\right),
\ee

\noindent
which by taking into account (\ref{II-18}) and after some algebraic manipulations can be put into the final form

\be
\label{II-32}
i\,g\,(2 \pi)^d\,\delta(p_1 + p_2 + p_3)\,\cos \left(p_1 \wedge p_2\right)\,.
\ee

\noindent
3) The interaction Lagrangian is taken to be

\be
\label{II-33}
{\cal L}_I\,=\,\frac{g}{4!}\,\phi(x) \ast \phi(x)\ast \phi(x)\,\ast \phi(x).
\ee

\noindent
In turns, Wick's theorem leads to

\bea
\label{II-34}
&&\int\,\left(\prod_{j = 1}^{4} dx_j\right)\,e^{i p^1_{\mu} x_1^{\mu} + i p^2_{\mu} x_2^{\mu} + i p^3_{\mu} x_3^{\mu}\,+\,i p^4_{\mu} x_4^{\mu}}\,<0|T(\phi(x_1)  \phi(x_2) \phi(x_3) \phi(x_4)\,\hat{S}_1)|0>\nonumber\\
&=&\,i \,\frac{g}{4!}\,\tilde{G}(p_1)\,\tilde{G}(p_2)\,\tilde{G}(p_3)\,\tilde{G}(p_4)
\nonumber\\
&\times&\,\int \,\left(\prod_{j = 1}^{4} dx_j\right)\,V(x_1, x_2, x_3, x_4)\,\sum_P\,e^{\left(i p^1_{\mu} x_{\alpha_1}^{\mu} + i p^2_{\mu} x_{\alpha_2}^{\mu} + i p^3_{\mu} x_{\alpha_3}^{\mu}\,+\,i p^4_{\mu} x_{\alpha_4}^{\mu}\right)}\,,
\eea

\noindent
where $\sum_P$ extends over all permutations of $\alpha_1$, $\alpha_2$, $\alpha_3$, and $\alpha_4$. Then, one goes through the usual steps: a) amputate the two-point functions, b) use Eq.(\ref{II-18}), and c) carry out appropriate algebraic rearrangements, to find

\bea
\label{II-35}
&&\frac{i g}{3}\,(2 \pi)^d\,\delta(p_1 + p_2 + p_3 + p_4)\,\left[\cos \left(p_1 \wedge p_2 + p_1 \wedge p_3 + p_2 \wedge p_3\right)\right.\nonumber\\
&&\left.+\,\cos \left(p_1 \wedge p_2 + p_1 \wedge p_3 - p_2 \wedge p_3\right)\,+\,\cos \left(p_1 \wedge p_2 - p_1 \wedge p_3 - p_2 \wedge p_3\right)\right]\,.
\eea

\noindent
The procedure for going from a given interaction to the corresponding vertex should by now be clear. For the next two bosonic interactions we restrict ourselves to quote the final form of the corresponding vertices.

\noindent
4) For the Lagrangian density

\be
\label{II-36}
{\cal L}_I\,=\,\frac{g}{4}\,\phi_1(x) \ast \phi_1(x)\ast \phi_2(x)\,\ast \phi_2(x)\,,
\ee

\noindent
the associated vertex reads

\be
\label{II-37}
i\,g\,(2 \pi)^d\,\delta(p_1 + p_2 + p_3 + p_4)\,\cos \left(p_1 \wedge p_2\right)\,\cos \left(p_3 \wedge p_4\right)\,.
\ee

\noindent
5) On the other hand, for

\be
\label{II-38}
{\cal L}_I\,=\,\frac{g}{4}\,\phi_1(x) \ast \phi_2(x)\ast \phi_1(x)\,\ast \phi_2(x)\,,
\ee

\noindent
one obtains as vertex

\be
\label{II-39}
i\,g\,(2 \pi)^d\,\delta(p_1 + p_2 + p_3 + p_4)\,\cos \left(p_1 \wedge p_2\,+\,p_3 \wedge p_4\right)\,.
\ee

The last two examples will be concerned with fermions.

\noindent
6) Let $\psi_1(x)$ and $\psi_2(x)$ be Dirac fermion fields and assume that

\be
\label{II-40}
{\cal L}_I\,=\,\frac{g}{4}\,\bar{\psi}_1(x) \ast \bar{\psi}_1(x) \ast \psi_2(x) \ast \psi_2(x)\,.
\ee

\noindent
To obtain the corresponding vertex we start by looking for the four-point connected Green function $<0|T\left(\psi_1(x_1) \psi_1(x_2) \bar{\psi}_2(x_3) \bar {\psi}_2(x_4) \hat{S}_1\right)|0>$, where the lowest order non-trivial contribution to the $\hat{S}$ operator is presently given by

\bea
\label{II-41}
\hat{S}_1\,&=&\,i\,\int d^dx\,{\cal L}_I\,=\,i\,\frac{g}{4}\,\int d^dx\,\bar{\psi}_1(x) \ast \bar{\psi}_1(x) \ast \psi_2(x) \ast \psi_2(x)
\nonumber\\
&=&\,i\,\frac{g}{4}\,\int\, d^dx_1\,d^dx_2\,d^dx_3\,d^dx_4\,V(x_1, x_2, x_3, x_4)\,\bar{\psi}_1(x_1) \bar{\psi}_1(x_2)  \psi_2(x_3)  \psi_2(x_4)\,,
\eea

\noindent
where Eq.(\ref{II-17}) has again been invoked but now involving fermionic fields. As in the bosonic case, we employ Wick's theorem and obtain

\bea
\label{II-42}
&&\int \left( \prod_{j = 1}^{4} dx_j\right)\,e^{i p^1_{\mu} x_1^{\mu} + i p^2_{\mu} x_2^{\mu} + i p^3_{\mu} x_3^{\mu}+ i p^4_{\mu} x_4^{\mu}}\,<0|T\left(\psi_1(x_1) \psi_1(x_2) \bar{\psi}_2(x_3) \bar {\psi}_2(x_4) \hat{S}_1\right)|0>\nonumber\\
&&= i\, \frac{g}{4}\,S_1(p_1)\,S_1(p_2)\,S_2(- p_3)\,S_2(- p_4)\int \left( \prod_{j = 1}^{4} dx_j\right)\,V(x_1, x_2, x_3, x_4)\,\left(e^{i p^1_{\mu} x_1^{\mu} + i p^2_{\mu} x_2^{\mu} + i p^3_{\mu} x_3^{\mu}+ i p^4_{\mu} x_4^{\mu}}\right.\nonumber\\
&&\left.-\,e^{i p^1_{\mu} x_1^{\mu} + i p^2_{\mu} x_2^{\mu} + i p^3_{\mu} x_4^{\mu} + i p^4_{\mu} x_3^{\mu}}\,-\,e^{i p^1_{\mu} x_2^{\mu} + i p^2_{\mu} x_1^{\mu} + i p^3_{\mu} x_3^{\mu}+ i p^4_{\mu} x_4^{\mu}}\,+\,e^{i p^1_{\mu} x_2^{\mu} + i p^2_{\mu} x_1^{\mu} + i p^3_{\mu} x_4^{\mu}+ i p^4_{\mu} x_3^{\mu}}\right)\,,
\eea

\noindent
where $S_1(p)$ and $S_2(p)$ denote the free propagators, in momentum space, of the respective fermionic species. After using Eq.(\ref{II-18}) one finds for the truncated four-point connected Green function the expression

\be
\label{II-43}
-\,i\,g\,(2 \pi)^d\,\delta(p_1 + p_2 + p_3 + p_4)\,\sin(p_1 \wedge p_2)\,\sin(p_3 \wedge p_4)\,.
\ee 

\noindent
7) We left as an exercise for the reader to show that the vertex associated with the interaction Lagrangian

\be
\label{II-44}
{\cal L}_I\,=\,\frac{g}{4}\,\bar{\psi}_1(x) \ast \psi_2(x) \ast \bar{\psi}_1(x) \ast \psi_2(x)\,,
\ee

\noindent
is 

\be
\label{II-45}
g\,(2 \pi)^d\,\delta(p_1 + p_2 + p_3 + p_4)\,\sin(p_1 \wedge p_2\,+\,p_3 \wedge p_4)\,.
\ee 

\section{THE UV/IR mechanism}
\label{sec:levelVII}

As a testing example of our previous calculations and with the purpose of illustrating about the UV/IR mechanism, we shall study in this Section some aspects of the $\phi^4$ theory formulated in a noncommutative four-dimensional space. The corresponding Lagrangian reads

\be
\label{II-46}
{\cal L}\,=\,\frac{1}{2}\, \pa_{\mu} \phi \,\pa^{\mu} \phi\,-\,\frac{1}{2}\,m^2\,\phi^2\,+\,{\cal L}_I(\phi)\,,
\ee

\noindent
where (observe the change of $g$ by $-g$ with respect to (\ref{II-33}))

\be
\label{II-47}
{\cal L}_I (\phi)\,=\,-\,\frac{g}{4!}\,\phi(x) \ast \phi(x)\ast \phi(x)\,\ast \phi(x).
\ee

\noindent
The Feynman rules for this model are:

\noindent
a) scalar field propagator $\Delta_F(p)$:

\be
\label{II-48}
\Delta_F(p)\,=\,\frac{i}{p^2 - m^2 +i\e}\,,
\ee

\noindent
b) quartic vertex (obtained from (\ref{II-35}) after replacing $g$ by $- g$):

\bea
\label{II-49}
&&\frac{- i g}{3}\,(2 \pi)^4 \,\delta(p_1 + p_2 + p_3 + p_4)\,\left[\cos \left(p_1 \wedge p_2 + p_1 \wedge p_3 + p_2 \wedge p_3\right)\right.\nonumber\\
&&\left.+\,\cos \left(p_1 \wedge p_2 + p_1 \wedge p_3 - p_2 \wedge p_3\right)\,+\,\cos \left(p_1 \wedge p_2 - p_1 \wedge p_3 - p_2 \wedge p_3\right)\right]\,.
\eea

We shall be looking for the lowest order perturbative correction ($\Sigma(p)$) to the two-point one particle irreducible (1PI) function ($\tilde{\Gamma}^{(2)}(p)$). As is well known,

\be
\label{II-50}
\tilde{\Gamma}^{(2)}(p)\,=\,p^2 - m^2 - \Sigma(p)\,,
\ee

\noindent
where, up to the order $g$, $ - i \Sigma(p)$ is only contributed by the tadpole diagram depicted in Fig.1, i.e.,

\be
\label{II-51}
- i \Sigma(p)\,=\,-\frac{i g}{3}\,\int\,\frac{d^4k}{(2 \pi)^4}\,\frac{i}{k^2 - m^2 + i\e}\,-\,\frac{i g}{6}\, \int\,\frac{d^4k}{(2 \pi)^4}\,\frac{i}{k^2 - m^2 + i\e}\,e^{2i k\wedge p}\,.
\ee

\noindent
For arriving at Eq.(\ref{II-51}) we first eliminated from (\ref{II-49}) the overall factor $(2 \pi)^4 \,\delta(p_1 + p_2 + p_3 + p_4)$  and, then, took into account that for the one loop correction to the two-point function the correct combinatoric factor is $4 . 3/4! = 1/2$ instead of one\cite{Ramond1}. The momenta were chosen as indicated in Fig.1, i.e., $p_3 = -p_4 = k$ and $ p_1 = - p_2 = p$. Correspondingly, the bracket in Eq.(\ref{II-49}) collapses into

\be
\label{II-52}
\left[ 2 \, +\, \cos\left(2 k \wedge p\right)\right].
\ee   

\noindent
Finally, parity arguments allowed the replacement of $\cos\left(2 k \wedge p\right)$ by $\exp(2 i k \wedge p)$. 

The first term in the right hand side of Eq.(\ref{II-51}) is the so called planar contribution ($ - i \Sigma(p)_P$ )and its analytic expression is, up to a numerical factor, that of $ - i \Sigma(p)$ in the commutative case\footnote{In the more general case a planar graph may contain a phase factor depending on the external momenta}. The second term, to be referred to as the nonplanar contribution ($ - i \Sigma(p)_{NP}$ ), contains an oscillatory factor which improves its UV behavior. The reason for the designation of a Feynman graph as being planar or nonplanar is easily understood if one uses the double line notation. If there are no crossing of lines the graph is said to be planar and nonplanar otherwise\cite{Minwalla1}. 

Power counting tell us that $\Sigma(p)_P$ is quadratically divergent. By using dimensional regularization one obtains

\be
\label{II-53}
\Sigma(p)_P\,=\,-\,\frac{g m^2}{48 \pi^2}\,\left[\frac{1}{\e}\,+\,\psi(2)\,-\,\ln \left(\frac{m^2}{ 4 \pi \mu^2}\right)\,+\,O(\e)\right]\,.
\ee

\noindent
Here, $\e \equiv 2 - d/2$, $\mu$ is the mass scale introduced by the regularization and $\psi(x) = d\Gamma(x)/dx $. As for the nonplanar part we find the finite result

\be
\label{II-54}
\Sigma(p)_{NP}\,=\,\frac{g\,m^2}{24 \pi^2}\,\sqrt{\frac{1}{m^2\,p \circ p}}\,K_1\left(\sqrt{m^2\,p \circ p}\right)\,,
\ee

\noindent 
where $K_1$ is the modified Bessel function and

\be
\label{II-55}
p \circ p\,\equiv\,p_{\mu}\,\left(\Theta^2\right)^{\mu \nu}\,p_{\nu}\,.
\ee

\noindent
By going back with (\ref{II-53}) and (\ref{II-54}) into (\ref{II-50}) and after mass renormalization we end up with

\be
\label{II-56}
\tilde{\Gamma}_R^{(2)}(p)\,=\,p^2 - m_R^2 - \,\frac{g\,m_R^2}{24 \pi^2}\,\sqrt{\frac{1}{m_R^2\,p \circ p}}\,K_1\left(\sqrt{m_R^2\,p \circ p}\right)\,,
\ee

\noindent
where $\tilde{\Gamma}_R^{(2)}(p)$ denotes the corresponding renormalized 1PI two-point function. One can verify that the renormalized mass ($m_R$) is given in terms of the unrenormalized mass ($m$) as follows

\be
\label{II-57}
m_R^2\,=\,m^2\,\left\{1\,-\,\frac{g}{48 \pi^2}\, \left[\frac{1}{\e}\,+\,\psi(2)\,-\,\ln \left(\frac{m^2}{ 4 \pi \mu^2}\right)\right]\right\}\,.
\ee

It remains to be analized the infrared behavior of $\tilde{\Gamma}_R^{(2)}(p)$. As is known\cite{Gradshteyn1}

\be
\label{II-58}
\lim_{p \circ p \rightarrow 0}\sqrt{\frac{1}{m_R^2\,p \circ p}}\,K_1\left(\sqrt{m_R^2\,p \circ p}\right)\,\longrightarrow\,\frac{1}{m_R^2\,p \circ p}\,+\,\frac{1}{2}\,\ln\left(\frac{\sqrt{m_R^2\,p \circ p}}{2}\right)\,,
\ee

\noindent
revealing the presence of quadratic and logarithmic infrared divergences. To summarize, in the commutative situation $\Sigma(p)$ is UV divergent but is not afflicted by IR singularities. On the other hand, if noncommutativity is present a piece of $\Sigma(p)$ becomes IR divergent, irrespective of the fact that the theory only contains massive excitations. This is the UV/IR mixing mechanism\cite{Minwalla1} to which we make reference in the opening paragraphs of the present Chapter. The insertion of $\Sigma(p)$ in higher order loops, as indicated in Fig.2, produces a harmful IR singularity which may invalidate the perturbative expansion\cite{Minwalla1}. The noncommutative $\phi^4$ model has been analyzed in Ref.\cite{Arefeva1} and shown to be renromalizable up to two-loops in spite of the presence of UV/IR mixing. This has been extended to all orders of perturbation theory within the context of the Wilsonian renormalization group\cite{Griguolo1} in which a cutoff is introduced and no IR singularities appear. 

A good estrategy in seeking for renormalizable noncommutative field theories is to look for models exhibiting, at most, logarithmic UV divergences in their commutative counterparts; then, the UV/IR mixing mechanism only produces harmless IR singularities. This is the case for $\phi^3$ in four dimensions\cite{Gomes1} and, as we shall see in the forthcoming chapter, also for the noncommutative supersymmetric Wess-Zumino model in four dimensional space-time\cite{Girotti2,Buchbinder1}. In three dimensional space-time we are aware of at least two noncommutative renormalizable models: the supersymmetric $O(N)$ nonlinear sigma model\cite{Girotti4} and the $O(N)$ supersymmetric linear sigma model in the limit $N \rightarrow \infty$\cite{Girotti5}. For nonsupersymmetric gauge theories the UV/IR mechanism breaks down the perturbative expansion\cite{Matusis1,Ruiz1,Hayakawa1,Jabbari1,Bonora1,Armoni1,Ferrara1,Guralnik1}. One may, however, entertain the hope that supersymmetric gauge theories are still renormalizable and free of nonintegrable infrared singularities\cite{Matusis1,Zanon1,Zanon2,Bichl1,Girotti6}

We close this chapter with a few comments concerning unitarity, causality, symmetries and the spin-statistic connection in noncommutative quantum field theories. 

For simplicity, the functional approach is the preferred framework for quantizing noncommutative field theories. It has been shown, within this approach, that field theories with space-time noncommutativity ($\Theta^{0i} \neq 0$) do not have a unitary S-matrix\cite{Gomis1}. However, recently\cite{Bahns1} it was claimed that the just mentioned violation of unitarity reflects an improper definition of chronological products. This may well be the case since quantum field theory on the noncommutative space-time is equivalent to a nonlocal theory. Owing to the nonlocality the various equivalent formulations of quantum field theories on Minkowski space are not longer equivalent on noncommutative spaces.

Furthermore, nonvanishing space-time noncommutativity also leads to a violation of causality, as shown in Ref.\cite{Seiberg2} and confirmed in Ref.\cite{Girotti3} for the noncommutative supersymmetric four-dimensional Wess-Zumino model.

Last but not least, the presence of a constant matrix ($\Theta^{\mu \nu}$) breaks Lorentz invariance as well as the discrete symmetries, parity (${\hat P}$), time reversal (${\hat T}$) and charge conjugation (${\hat C}$), although $ {\hat P}{\hat C}{\hat T}$ symmetry is preserved irrespective of the form of $\Theta^{\mu \nu}$\cite{Chaichian1,Jabbari2}. In the case of only space-space noncommutativity ($\Theta^{0i} = 0$) the parity of a noncommutative field theory is the same as for its commutative counterpart but time reversal and charge conjugation are broken.

As for the spin-statistics theorem, it holds for theories with space-space noncommutativity. Up to our knowledge, no definite statement can be made in the cases of space-time and light-like ($ \Theta^{\mu \nu}\Theta_{\mu \nu} = 0$) noncommutativity\cite{Chaichian1}. 
            
\newpage

\begin{center}
{\bf Chapter III: THE NONCOMMUTATIVE WESS-ZUMINO MODEL}
\end{center}

\section{Introduction}
\label{sec:level8}

This Chapter is mainly based in Refs.\cite{Girotti2,Girotti3}. We shall first show that the noncommutative Wess-Zumino model in four dimensions is a consistent quantum field theory in the sense of being ultraviolet renormalizable and free of the dangerous UV/IR mixing at any arbitrary order of perturbation theory. Afterwards, the non-local effects produced by the noncommutativity are illustrated by studying the low energy behavior of the Wess-Zumino model. 

In four dimensional Minkowski space-time the Wess-Zumino model is defined by the Lagrangian density\cite{Wess1}

\bea
\label{III-1}
{\cal L} &= &\frac12  A(-\partial^2)A+\frac12  B(-\partial^2)B
+ \frac12 \overline \psi(i\not \! \partial - m)\psi +\frac12 F^2+\frac12 G^2
+ m F A + m G B +\nonumber \\
&\phantom a & g (F A^2 -  F B^2 + 2 G A B -  \overline \psi
\psi A - i  \overline \psi \gamma_5\psi B )\,,
\eea

\noindent
where $A$ is a scalar field, $B$ is a pseudo scalar field, $\psi$ is
a Majorana spinor field and $F$ and $G$ are, respectively, scalar and
pseudoscalar auxiliary fields\footnote{ Our Minkowskian metric is $g^{00} = - g^{11} = - g^{22} = - g^{33} = + 1$. Furthermore, we use Dirac's representation for the $\gamma$ matrices and $\gamma_5 \equiv i \gamma^0 \gamma^1 \gamma^2 \gamma^3$ implying that $\gamma_5^{\dagger} = \gamma_5$ and $\gamma_5^2 = 1$.}. By extending the above model to a
noncommutative space one is led to the Lagrangian density

\be
\label{III-2}
{\cal L} ={\cal L}_0\,+\,{\cal L}_m\,+\,{\cal L}_g\,,
\ee

\noindent
where

\bml
\label{III-211}
\bea
&& {\cal L}_0\,=\,\frac12  A(-\partial^2)A+\frac12  B(-\partial^2)B
+ i \frac12 \overline \psi \not \! \partial \psi +\frac12 F^2+\frac12 G^2\,,\label{mlett:aIII-211}\\
&& {\cal L}_m\,=\, m F A + m G B - \frac{m}{2} \overline \psi \psi\,,\label{mlett:bIII-211}\\ 
&& {\cal L}_g\,=\, g (F\star A \star
A- F\star B \star B + G\star A \star B + G \star B \star A -\overline \psi
\star \psi\star A - \overline \psi\star  i\gamma_5 \psi \star B)\,.\label{mlett:cIII-211}
\eea
\eml

\noindent
The Lagrangian (\ref{III-2}) was also written using the superspace formalism in
\cite{Ferrara1,Terashima1}. However, we will work with components fields in
order to trace the effects of noncommutativity in the divergent 
Feynman integrals. In order to evade causality and unitarity problems\cite{Gomis1} and to preserve parity\cite{Chaichian1} we shall assume from now on that $\Theta_{0i} = 0$. 

It should also be noticed that there is only one
possible extension of the cubic term $2 G A B$, to the noncommutative
case, which preserves supersymmetry. The elimination of the auxiliary fields through their corresponding equations of motion turns the bilinear terms in the Lagrangian
Eq.(\ref{III-2}) into the standard mass terms. On the other hand, the cubic terms
produce quartic interactions which, in terms of a complex field
$\phi=A+iB$, can be cast as $\phi^* \star \phi^* \star \phi \star
\phi$. This potential belongs to a class of non-renormalizable
potentials, as discussed in \cite{Arefeva2}. As it will be
shown below, supersymmetry saves the day turning the theory into a
renormalizable one. 

The propagators for the $A$ and $F$ fields are (see Fig.3)

\bml
\label{III-3}
\bea
\Delta_{AA}(p) &=& \Delta(p)\equiv\frac{i}{p^2-m^2+i\epsilon}\,,\label{mlett:aIII-3}\\
\Delta_{FF}(p) &=& p^2 \Delta(p)\,,\label{mlett:bIII-3}\\
\Delta_{AF}(p) &=&\Delta_{FA}(p) = -m \Delta(p)\,,\label{mlett:cIII-3}
\eea
\eml

\noindent
whereas the propagators involving the $B$ and $G$ fields have identical
expression (i.e., they are obtained by replacing $A$ by $B$ and $F$ by $G$).
For the $\psi$ field we have

\be
\label{III-4}
S(p)= \frac{i}{\not \! p -m} \label{4}\,.
\ee

The analytical expressions associated to the vertices are:

\bml
\label{III-5}
\bea
F A^2 \quad {\mbox {vextex:}}&& \quad ig \cos(p_1\wedge p_2)\,,\label{mlett:aIII-5} \\
F B^2  \quad {\mbox {vextex:}}&& \quad -ig \cos(p_1\wedge p_2)\,,\label{mlett:bIII-5}\\
G A B \quad  {\mbox {vertex:}} && \quad 2 ig \cos (p_1\wedge p_2)\,,\label{mlett:cIII-5}\\
\overline \psi \psi A\quad {\mbox {vertex:}} &&\quad -ig 
\cos (p_1\wedge p_2)\,,\label{mlett:dIII-5}\\ 
\overline \psi \psi B\quad {\mbox {vertex:}} &&\quad -ig\gamma_5 
\cos (p_1\wedge p_2)\,.\label{mlett:eIII-5}
\eea
\eml

Due to the oscillating factors provided by the cosines some of the
integrals constructed with the above rules will be finite but in general
divergences will survive, the degree of superficial divergence
for a generic 1PI graph $\gamma$ being\footnote{We shall designate by ${\cal O}$ either a generical field in the model or the collection of all fields. The role played by this symbol should be clear from the corresponding mathematical expression.}   

\be
\label{III-6}
d(\gamma)= 4 -  I_{AF} -I_{BF}-N_A-N_B-2 N_F-2N_G - \frac32 N_\psi\label{5}\,,
\ee

\noindent
where $N_{\cal O}$ denotes the number of external lines associated to the 
field ${\cal O}$ and $ I_{AF}$ and $I_{BF}$ are the numbers of internal
lines associated to the indicated mixed propagators.  

\section{Ward identities}
\label{sec:level9}

By following the strategy developed in Ref.\cite{Iliopoulos1}, to prove the renormalizability of the commutative Wess-Zumino model, we first look for the Ward identities deriving from the fact that ${\cal L}_0$, ${\cal L}_m$, and ${\cal L}_g$, which enter in the action

\be
\label{III-7}
{\cal S}[{\cal O}] \,\equiv\,\int d^4x\,{\cal L ({\cal O})}\,,
\ee

\noindent
are separately invariant under the supersymmetry transformation

\bml
\label{III-8}
\bea
&&\delta A\,=\,\bar{\alpha}\,\psi\,,\label{mlett:aIII-7}\\
&&\delta B\,=\,-\,i\,\bar{\alpha}\,\gamma_5\,\psi\,,\label{mlett:bIII-7}\\
&&\delta \psi\,=\,-\,i\,\not \!\partial (A - i \gamma_5 B)\,\alpha\,+\,(F - i \gamma_5\,G)\,\alpha\,,\label{mlett:cIII-7}\\
&&\delta F\,=\,-\,i\,\bar{\alpha}\,\not \!\partial \psi\,,\label{mlett:dIII-7}\\
&&\delta G\,=\,-\,\bar{\alpha}\,\gamma_5\,\not \!\partial \psi\,,\label{mlett:eIII-7}
\eea
\eml

\noindent
where $\alpha$ is a constant totally anticommuting Majorana spinor. Since supersymmetry transformations are linear in the fields they are not affected by the replacement of ordinary field products by Moyal ones. Moreover, the proof of invariance of ${\cal S}[{\cal O}]$ under (\ref{III-8}) is based on the observation that

\be
\label{III-9}
\delta \left(A \ast B\right)\,=\,\delta A \ast B\,+\,A \ast \delta B\,.
\ee

We introduce next external sources for all fields ($J_{{\cal O}}$). In the presence of external sources the action is modified as follows

\be
\label{III-10}
{\cal S}[{\cal O} , J_{\cal O}]\,=\,{\cal S}[{\cal O}]\,+\,\int d^4x \,\left(J_A\,A\,+\,J_B\,B\,+\,J_F\,F\,+\,J_G\,G\,+\,\bar{\eta} \psi \right)\,.
\ee

\noindent
We shall designate by $Z[J_{\cal{O}}]$ the generating functional of Green functions, i.e.,

\be
\label{III-11}
Z[J_{\cal{O}}]\,=\,\int\,\left[D{\cal{O}}\right]\,e^{{\cal S}[{\cal O} , J_{\cal O}]}\,.
\ee

\noindent
The generating functional $Z[J_{\cal{O}}]$ is insensitive to a change of dummy integration variables in the right hand side of (\ref{III-11}). In particular, we are interested in ${\cal O} \rightarrow  {\cal O} + \delta {\cal O}$ with $\delta {\cal O}$ given by (\ref{III-8}). This change of integration variables leaves $S[{\cal O}]$ invariant and its corresponding Jacobian is just a constant. By retaining terms up to the first order in $\delta {\cal O}$ one finds that

\be
\label{III-12}
\int\,\left[D{\cal{O}}\right]\,e^{{\cal S}[{\cal O} , J_{\cal O}]}\,\int\,d^4x\,\sum_{{\cal O}}\,J_{\cal O} \,\delta {\cal O}\,=\,0\,.
\ee

\noindent
Let us assume next that the external sources transform according to

\bml
\label{III-13}
\bea
&& \delta J_A\, = \,-i\,\left(\pa_{\mu} \bar{\eta}\right)\,\gamma^{\mu}\,\alpha\,,\label{mlett:aIII-13}\\ 
&& \delta J_B\,=\,-\,\left(\pa_{\mu} \bar{\eta}\right)\,\gamma^{\mu}\,\gamma_5\,\alpha\,,\label{mlett:bIII-13}\\ 
&& \delta J_F\,=\,- \bar {\eta} \alpha\,,\label{mlett:cIII-13}\\
&& \delta J_G\,=\,i\, \bar{\eta}\,\gamma_5\, \alpha\,,\label{mlett:dIII-13}\\
&& \delta \eta\,=\,\pa_{\mu} \left( i J_F - \gamma_5 J_G \right)\,\gamma^{\mu}\,\alpha\,-\,\left( J_A - i \gamma_5 J_B\right) \alpha\,.\label{mlett:eIII-13}
\eea
\eml

\noindent
It is an easy task to verify that the simultaneous transformation of fields and sources according to (\ref{III-8}) and (\ref{III-13}), respectively, leaves ${\cal S}[{\cal O} , J_{\cal O}]$ invariant. In other words, it amounts to

\be
\label{III-14}
\delta \left(  \sum_{{\cal O}}\,J_{\cal O} \,{\cal O}\right)\,=\,0\,.
\ee

\noindent
By going back with (\ref{III-14}) into (\ref{III-12}) one obtains

\be
\label{III-15}
-\,\int\,\left[D{\cal{O}}\right]\,e^{{\cal S}[{\cal O} , J_{\cal O}]}\,\int\,d^4x\,\sum_{{\cal O}}\,( \delta J_{\cal O}) \,{\cal O}\,=\,0\,,
\ee

\noindent
or, equivalently,

\bea
\label{III-16}
&&\frac{\delta Z}{\delta J_A}\,(\pa_{\mu} \bar{\eta})\,\gamma^{\mu}\,-\,i\,\frac{\delta Z}{\delta J_B}\,(\pa_{\mu} \bar{\eta})\,\gamma^{\mu}\,\gamma_5\,-\,i\,\frac{\delta Z}{\delta J_F}\,\bar{\eta}\,-\,\frac{\delta Z}{\delta J_G}\,\bar{\eta}\,\gamma_5\nonumber\\
&&+\,i\,\frac{\delta Z}{\delta \eta}\,\left[\gamma^{\mu}\,\pa_{\mu} \left(i J_F + \gamma_5 J_G \right)\,-\,\left( J_A - i \gamma_5 J_B\right) \right]\,=\,0\,,
\eea

\noindent
which gives origin to a family of Ward identities. We shall always
assume that UV divergent integrals have been regularized in such a way that these Ward identities are preserved. The precise form of the regularization is irrelevant, as far as it obeys the usual additive rules employed in the calculation of Feynman diagrams\footnote{ A regularization method that preserves supersymmetry, at least up to two loops, has been proposed in Ref.\cite{Iliopoulos1}.}.

It is more useful to express the Ward Identities in terms of one-particle irreducible Green functions (vertex functions). To this end we first introduce the generating functional of connected Green functions $W[J_{\cal O}]$ by means of

\be
\label{III-17}
W[J_{\cal O}]\,=\,-\,i\,\ln Z[J_{\cal O}]\,.
\ee

\noindent
Then, the generating functional of vertex functions ($\Gamma[R_{\cal O}]$) is defined as

\be
\label{III-18}
\Gamma[R_{\cal O}]\,\equiv\,W[J_{\cal O}]\,-\,\int d^4 x\,\left(J_A R_A\,+\,J_B R_B\,+\,J_F R_F\,+\,J_G R_G\,+\,\bar{\eta} R_{\psi}\right)\,,
\ee

\noindent
where

\bml
\label{III-19}
\bea
&& R_{\phi}(x)\,\equiv\,\frac{\delta W[J_{\cal O}]}{\delta J_{\phi}(x)}\,,\label{mlett:aIII-19}\\
&& R_{\psi}(x)\,\equiv\,\,\frac{\delta W[J_{\cal O}]}{\delta \bar{\eta}(x)}\,,\label{mlett:bIII-19}\\
&& J_{\phi}(x)\,\equiv\,-\,\frac{\delta \Gamma[R_{\cal O}]}{\delta R_{\phi}(x)}\,,\label{mlett:cIII-19}\\
&& \bar{\eta}(x)\,\equiv\,-\,\frac{\delta \Gamma[R_{\cal O}]}{\delta R_{\psi}(x)}\,.\label{mlett:dIII-19}
\eea
\eml

\noindent
Here, $R_{\phi}$ ($J_{\phi}$) designates the subsets of scalar and pseudoscalar fields (sources). By replacing (\ref{III-19}) into (\ref{III-16}) one arrives to

\bea
\label{III-20}
&& -\,i\,R_{\psi}\,\frac{\delta \Gamma}{\delta R_A}\,-\,\gamma_5\,R_{\psi}\,\frac{\delta \Gamma}{\delta R_B}\,-\,\gamma^{\mu}\,\left(\pa_{\mu} R_{\psi}\right)\,\frac{\delta \Gamma}{\delta R_F}
\,+\,i\,\gamma_5\,\gamma^{\mu}\,\left(\pa_{\mu} R_{\psi}\right)\,\frac{\delta \Gamma}{\delta R_G}
\nonumber\\
&&-\left(i R_F\,+\,\gamma_5\,R_G\,-\,\pa_{\mu}R_A\,\gamma^{\mu}\,+\,i\,\pa_{\mu}R_B\,\gamma_5\,\gamma^{\mu}\, \right)\,\frac{\delta \Gamma}{\delta {\bar R}_{\psi}}\,=\,0\,,
\eea

\noindent
which is the desired expression.

We introduce next regularization dependent supersymmetric invariant counterterms which make all Green functions finite, even after the regularization is removed. The renormalized theory is obtained by assigning prescribed values to the primitively divergent vertex functions at some subtraction point that we choose to be the origin in momentum space. It is a consequence of the Ward identities (\ref{III-16}) and (\ref{III-20}) that all the tadpoles vanish and, therefore, we do not need to introduce counterterms linear in the fields. 

On the other hand, power counting (see Eq.(\ref{III-6})) tells us that we shall need, in principle, ten parameters for the two-point functions, seven for the three-point and three for the four-point functions. Once arbitrary values are assigned to these quantities, we can remove the regulator and obtain a finite answer. The key point is, however, that this assignment can be done in agreement with the Ward identities and, hence, preserving the supersymmetry at the level of the renormalized quantities. For the details we refer the reader to Ref.\cite{Iliopoulos1}. The main outcomes are that no counterterms of the form $A^2$, $B^2$, $A^3$, $A B^2$, $A^4$, $A^2 B^2$ and $B^4$ are needed. At most a common wave function, coupling constant and mass renormalizations are required. We shall see in the next section how all this work at the one-loop level.  

\section{The one-loop approximation}
\label{sec:level10}

One may verify that, at the one-loop level, all the tadpoles contributions add up to zero. This confirms the outcomes obtained in the previous Section by exploring the Ward identities. 

Let us now examine the contributions to the self-energy of the $A$
field. The corresponding graphs are
those shown in Fig.4$a$-4$e$. In that figure diagrams $a$,$b$ and $ c$
are quadratically divergent whereas graphs $d$ and $e$ are
logarithmically divergent.  We shall first prove that the quadratic
divergences are canceled. In fact, we have that

\bea
\label{III-21}
\Gamma_{{\mbox 4}a-c}(AA)&=&-g^2\int \frac{d^4k}{(2\pi)^4 }\cos^2(k\wedge p) 
\{4 k^2+4 k^2-2
{\rm Tr}[(\not \!k+\not \!p+m)(\not \!k+m)]\}\nonumber \\
&\phantom a&\times  \Delta(k+p)  \Delta(k)\,,
\eea

\noindent
where the terms in curly brackets correspond to the graphs $a$, $b$ and $c$,
respectively. After calculating the trace we obtain

\be
\label{III-22}
\Gamma_{{\mbox 4}a-c}(AA)=8 g^2\int \frac{d^4k}{(2\pi)^4 }(p\cdot k + m^2)
\cos^2(k\wedge p) \Delta(k) \Delta(k+p)\,.
\ee

\noindent
This last integral is, at most, linearly divergent. However, the
would be linearly divergent term vanishes by symmetric integration
thus leaving us with an integral which is, at most, logarithmically
divergent.  Adding to Eq.(\ref{III-22}) the contribution of the graphs 2$d$
and 2$e$ one arrives at 

\be
\label{III-23}
\Gamma_{{\mbox 4}a-e}(AA)= 8 g^2\int \frac{d^4k}{(2\pi)^4 }
\cos^2(p\wedge k) (p\cdot k) \Delta(k) \Delta(k+p)\,.
\ee

\noindent 
To isolate the divergent contribution to $ \Gamma_{{\mbox 2}a-e}(AA)$
we Taylor expand the coefficient of $\cos^2(p\wedge k)$ with respect
to the variable $p$ around $p=0$, namely,

\bea
\label{III-24}
&& 8 g^2\int\frac{d^4k}{(2\pi)^4}\cos^2(p\wedge k) t^{(1)}(p)\left[ (p\cdot k) \Delta(k) \Delta(k+p)\right]\Bigr.|_{p=0}\nonumber \\
&=& 16 g^2\int\frac{d^4k}{(2\pi)^4}\cos^2(p\wedge k)\frac{(p \cdot k)^2}{(k^2- m^2)^3} \,,   
\eea

\noindent 
where $t^{(r)}(p)$ denotes the Taylor operator of order $r$. Since 
$\cos^2(k\wedge p)= (1+\cos(2k\wedge p))/2$ the divergent part of
(\ref{III-24}) is found to read

\be
\label{III-25}
\Gamma_{Div}(AA)= 2 g^2 p^2 \int \frac{d^4k}{(2\pi)^4 } \frac{1}{(k^2-m^2)^2}\equiv
i I_\xi g^2 p^2\,,
\ee

\noindent
where the subscript $\xi$ remind  us that all  integrals
are regularized through the procedure indicated in \cite{Iliopoulos1}.
In the commutative Wess-Zumino model this divergence occurs with a
weight twice of the above. As usual, it is 
eliminated by the wave function renormalization $A=Z^{1/2}A_r$,
where $A_r$ denotes the renormalized $A$ field. Indeed, it is easily
checked that with the choice $ Z=1- I_\xi g^2$ the contribution
(\ref{III-25}) is canceled.

We turn next into analyzing the term containing $\cos(2k\wedge p)$ in
(\ref{III-24}). For small values of $p$ it behaves as $p^2 \ln (p^2/m^2)$. Thus, in
contradistinction to the nonsupersymmetric $\phi^4_4$ case
\cite{Minwalla1}, there is no infrared pole and the function actually
vanishes at $p=0$.

One may  check that at one-loop the $B$ field self-energy is the
same as the self-energy for the $A$ field, i. e., $\Gamma(BB)=\Gamma(AA)$.
Therefore the divergent part of $\Gamma(BB)$ will be eliminated if
we perform the same wave function renormalization as we did for the
$A$ field, $B= Z^{1/2}B_r$. We also found that the mixed two point Green
functions do not have one-loop radiative corrections, $\Gamma(AF)=\Gamma(BG)=0$.

The one-loop corrections to the two point of the auxiliary field $F$
are depicted in Fig.5. The two graphs give identical contributions
leading to the result

\be
\label{III-26}
\Gamma(FF)= -4 g^2 \int \frac{d^4k}{(2\pi)^4 }\cos^2(k\wedge p) \Delta(k)
\Delta(k+p)\,,
\ee

\noindent 
whose divergent part is

\be
\label{III-27}
\Gamma_{Div}(FF)= 2 g^2 \int    \frac{d^4k}{(2\pi)^4 }\frac{1}{(k^2-m^2)^2}=
i I_\xi g^2\,,
\ee

\noindent
involving the same divergent integral of the two point functions of
the basic fields. It can be controlled by the field renormalization
$F= Z^{1/2} F_r$, as in the case of $A$ and $B$. Analogous reasoning
applied to the auxiliary field $G$ leads to the conclusion that $G
=Z^{1/2}G_r$.  However, things are different as far as the term
containing $\cos(2 k \wedge p)$ is concerned. It diverges as
$\ln(p^2/m^2)$ as $p$ goes to zero. Nevertheless, this is a harmless
singularity in the sense that its multiple insertions in higher order
diagrams do not produce the difficulties pointed out in
\cite{Minwalla1}.

Let us now consider the corrections to the self-energy of the spinor field
$\psi$ which are shown in Fig.6. The two contributing graphs give

\bea
\label{III-28}
\Gamma(\psi \overline \psi)&=& 4 g^2 \int \frac{d^4 k}{(2\pi)^4}
\cos^2(k\wedge p) \Delta(k)\Delta(k+p)[(\not \! k +m)- \gamma_5(\not\! k+m)\gamma_5]
\nonumber\\
& =& 8 g^2 \int \frac{d^4 k}{(2\pi)^4}
\cos^2(k\wedge p) \not \! k\, \Delta(k)\Delta(k+p)\,,
\eea

\noindent
so that for the divergent part we get $\Gamma_{Div}(\psi\overline
\psi)= ig^2\not\!\! p \,I_\xi$ leading to the conclusion that the
spinor field presents the same wave function renormalization of the
bosonic fields, i. e., $\psi = Z^{1/2} \psi_r$.  As for the term
containing $\cos(2k\wedge p)$ it behaves as $\not \! p
\ln(p^2/m^2)$ and therefore vanishes as $p$ goes to zero.

The one-loop superficially (logarithmically) divergent graphs
contributing to the three point function of the $A$ field are shown in
Fig.7. The sum of the amplitudes corresponding to the graphs
Fig.7$a$ and Fig.7$b$ is

\bea
\label{III-29}
\Gamma_{7a+7b}(AAA)&=& 96 ig^3 m \int\frac{d^4k}{(2\pi)^4} (k-p_2)^2 \Delta(k) \Delta(k+p_3)
\Delta(k-p_2)\cos( k\wedge p_1 + p_3\wedge p_1) \nonumber\\
&\phantom a &\times\cos(p_2\wedge k)\cos(p_3\wedge k)\,,
\eea

\noindent
while its divergent part is found to read

\be
\label{III-30}
\Gamma_{7a+7b\,\,Div}(AAA)= 24 ig^3 m \cos( p_3\wedge p_1) \int
\frac{d^4k}{(2\pi)^4} (k)^2 (\Delta(k))^3.
\ee

\noindent
The divergent part of the graph 7$c$, nonetheless, gives a similar
contribution but with a minus sign so that the two divergent parts add
up to zero. Thus, up to one-loop the three point function
$\Gamma(AAA)$ turns out to be finite. Notice that a nonvanishing
result would spoil the renormalizability of
the model. The analysis of $\Gamma(ABB)$ follows along similar lines
and with identical conclusions. Furthermore, it is not difficult to
convince oneself that $\Gamma(FAA)$, $\Gamma(FBB)$ and $\Gamma(GAB)$
are indeed finite. 

As for $\Gamma(A \psi\overline \psi)$ we notice that superficially
divergent contributions arise from the diagrams depicted in Figs.
8$a$ and 8$b$. In particular, diagram Fig.8$a$
yields

\bea
\label{III-31}
\Gamma_{8a}(A\psi\overline\psi)&=&8ig^3\int \frac{d^4k}{(2\pi)^4} \Delta(k) 
\Delta(p_2+k) \Delta (k-p_1)
(\not \! p_2+\not \! k+m)(\not \! k-\not \! p_1+m)\nonumber\\
&\phantom a&\times \cos(k\wedge p_3-p_1\wedge p_3) \cos(k\wedge p_1)\cos(k\wedge p_2)\,,
\eea

\noindent
while 8$b$ gives

\bea
\label{III-32}
\Gamma_{8b}(A\psi\overline\psi)&=&-8ig^3\int \frac{d^4k}{(2\pi)^4} \Delta(k) 
\Delta(p_2+k) \Delta( k-p_1)\gamma_5(\not \! p_2+\not \! k+m)
(\not \! k-\not \! p_1+m)\gamma_5\nonumber \\
&\phantom a&\times \cos(k\wedge p_3-p_1\wedge p_3) \cos(k\wedge p_1)
\cos(k\wedge p_2)\, , 
\eea

\noindent
so that the sum of the two contributions is also finite. The same
applies for $\Gamma(B\psi\overline\psi)$. 

We therefore arrive at another important result, namely, that there is
no vertex renormalization at the one loop level. This parallels the
result of the commutative Wess-Zumino model. 

To complete the one-loop analysis we must examine the four point
functions. Some of the divergent diagrams contributing to
$\Gamma(AAAA)$ are depicted in Fig.9$a-c$.  The analytical expression
associated with the graph Fig.9$a$ is

\bea
\label{III-33}
\Gamma_{9a}(AAAA) &=& 16g^4\int \frac{d^4k}{(2\pi)^4}k^2\Delta(k)\Delta (k+p_1)
(k+p_1+p_3)^2 \Delta(k+p_1+p_3)\Delta (p_2 -k)\nonumber \\
&\phantom a & \times\cos(k\wedge p_1) \cos(k\wedge p_2) 
\cos [(k+p_1)\wedge p_3]
\cos[(k-p_2)\wedge p_4].
\eea

\noindent
There are five more diagrams of this type, which are obtained by
permuting the external momenta $p_2\,\, , p_3$ and $p_4$ while keeping
$p_1$ fixed. Since we are interested in the (logarithmic) divergence
associated with this diagram, we set all the external momenta
to zero in the propagators but not in the arguments of the cosines. This 
yields

\bea
\label{III-34}
\Gamma_{9a\,\, Div}(AAAA) &=& 16g^4\int \frac{d^4k}{(2\pi)^4}(k^2)^2
(\Delta(k))^4\nonumber \\
&\phantom a & \times
\cos(k\wedge p_1) \cos(k\wedge p_2) \cos [(k+p_1)\wedge p_3]
\cos[(k-p_2)\wedge p_4]\,.
\eea

\noindent
Adopting the same procedure for the other five graphs we notice 
that the corresponding contributions are pairwise equal. The final
result is therefore

\bea
\label{III-35}
&&\Gamma_{Div}(AAAA) = 32g^4\int \frac{d^4k}{(2\pi)^4}(k^2)^2
(\Delta(k))^4  \cos(k\wedge p_1)\nonumber \\
&\phantom a &\times [ \cos(k\wedge p_2) \cos [(k+p_1)\wedge p_3]
\cos[(k-p_2)\wedge p_4] +p_3\leftrightarrow p_4+p_2\leftrightarrow p_4]\,.
\eea

There is another group of six diagrams, Fig.9$b$, which are obtained 
from the preceding
ones by replacing the propagators of $A$ and $F$ fields by the propagator of 
the $B$ and $G$ fields, respectively. The net effect of adding these
contributions is, therefore, just to double the numerical factor in the right
hand side of the above formula.

Besides the two groups of graphs just mentioned, there are another six
graphs with internal fermionic lines. A representative of this group
has been drawn in Fig.9$c$. It is straightforward to verify that
because of the additional minus sign due to the fermionic loop, there
is a complete cancellation with the other contributions described
previously.  The other four point functions may be analyzed similarly
with the same result that no quartic counterterms are needed.

\section{Absence of mass and coupling constant renormalization to all orders of perturbation theory}
\label{sec:level11}

In the previous section we proved that up to one loop the
noncommutative Wess-Zumino model is renormalizable and only requires a
common wave function renormalization. Here, we shall
prove that no mass and coupling constant counterterms are needed at
any finite order of perturbation theory. As in the commutative case,
our proof relies heavily on the Ward identities. 

After some straightforward algebra one can show that

\be
\label{III-36}
\int d^4 y\frac{\delta}{\delta {\cal O}(y)}\int d^4 x\,
\underbrace{{\cal O}(x)\star{\cal O}(x)\star\ldots \star{\cal O}(x)}_{n \rm 
\; factors}= n \int d^4 x\,\underbrace{{\cal O}(x)\star{\cal O}(x)\star
\ldots \star{\cal O}(x)}_{n-1 \rm \; factors}\,,
\ee

\noindent
which, in turns, enables one to find

\be
\label{III-37}
\int\,d^4y\,\frac{\delta}{\delta A(y)}\,\int\,d^4x\,{\cal L}_g\,=\,g\,\int\,d^4y\,\left[2\,F(y) A(y)\,+\,2\,G(y)\,B(y)2\,-\,\bar{\psi}(y) \psi(y)\right]\,,
\ee

\noindent
where (\ref{mlett:cIII-211}) has been used. Therefore, 

\be
\label{III-38}
\frac{\pa}{\pa m}\,\int\,d^4x\,{\cal L}_m(x)\,=\,\frac{1}{2g}\,\int\,d^4y\,\frac{\delta}{\delta A(y)}\,,
\ee

\noindent
as can be seen by using (\ref{mlett:bIII-211}). At the level of the generating functional $W[J_{\cal O}]$, this, together with $<F> = 0$, implies that

\be
\label{III-39}
\frac{\partial \phantom a}{\partial m} W[J_{\cal O}]= -\frac{m}{2 g}\int \frac{\delta
W[J_{\cal O}]}{\delta J_F(y)} d^4 y - \frac{iW[J_{\cal O}]}{2g} \int J_A(y) d^4 y\,,
\ee

\noindent
which looks formally identical to the corresponding relation in the commutative case\cite{Iliopoulos1}. We emphasize that the relationship above holds for unrenormalized but regularized connected Green functions.

In terms of the 1PI generating functional $\Gamma[R_{\cal O}]$ the identity
(\ref{III-39}) becomes 

\be
\label{III-40}
\frac{\partial}{\partial m}\Gamma[R_{\cal O}]= - \frac{m}{2g}\int R_F(y)d^4 y + 
\frac{1}{2g}\int\frac{\delta\Gamma[R_{\cal O}]}{\delta R_A(y)} d^4 y\,. 
\end{equation}

\noindent 
By taking the functional derivative with respect to $R_F$ and then
putting all $R$'s equal to zero we obtain

\be
\label{III-41}
m= \Gamma(FA)\Bigl |_{p^2=0}= Z^{-1}\Gamma_r(FA) \Bigl |_{p^2=0}\,,
\ee

\noindent
where $\Gamma_r(AF)$ is the renormalized 1PI Green function of the
indicated fields. We take as normalization conditions those specified
in \cite{Iliopoulos1}. Specifically, $\Gamma_r(FA) \Bigl
|_{p^2=0}=m_r$, where $m_r$ is taken to be the renormalized
mass. Hence, $m_r=Z m$ implying that there is no additive mass
renormalization. Through similar steps one also finds that
$g_r=Z^{3/2}g$, where $g_r$ is the renormalized coupling
constant. This implies the absence of coupling constant counterterms.

We stress the fact that, by exploiting the Ward identities, we have
succeeded in generalizing to all orders of perturbation theory the one
loop result concerned with the absence of counterterms different from those
already present in the original Lagrangian.

\section{The low energy limit of the noncommutative Wess-Zumino model}
\label{sec:level12}

Noncommutative field theories present many unusual properties. Their non-local character gives rise to to a mixing of UV and IR divergences which may spoil the renormalizability of the model. The only four dimensional noncommutative field theory known at present is the Wess-Zumino model. Hence, we have at our disposal an appropriate model for studying the non-local effects produced by the noncommutativity. To carry out this study we shall consider the NC Wess-Zumino model
and determine, at the tree level, the non-relativistic potentials mediating the fermion-fermion and boson-boson scattering along the lines of\cite{Sakurai1,Girotti7}.

We first concentrate on the elastic scattering of two Majorana
fermions. We shall designate by $p_1, p_2$ ($p'_1, p'_2$) and by
$\epsilon_1, \epsilon_2$ ($\epsilon'_1, \epsilon'_2$) the four momenta
and z-spin components of the incoming (outgoing) particles, respectively. The
Feynman graphs contributing to this process, in the lowest order of
perturbation theory, are those depicted in Fig.10\footnote{In these
diagrams the arrows indicate the flow of fermion number rather than
momentum flow.} while the associated amplitude is given
by $R = -i(2\pi)^4 \delta^{(4)}(p'_1 + p'_2 - p_1 - p_2)\, T$, where $ T = T_a +
T_b + T_c$ and

\bml
\label{III-42}
\bea
T_a\,&=&\, K\, \cos(p'_1 \wedge p_1) \, \cos(p'_2 \wedge p_2)\,\frac{\left(F_a - F_a^5\right)}{{\cal D}_a}\,, \label{mlett:aIII-42}\\
T_b\,&=&\,-\, K\, \cos(p'_1 \wedge p_2) \, \cos(p'_2 \wedge p_1)\,\frac{\left(F_b - F_b^5\right)}{{\cal D}_b}\,, \label{mlett:bIII-42}\\
T_c\,&=&\, K\, \cos(p'_1 \wedge p'_2) \, \cos(p_1 \wedge p_2)\,\frac{\left(F_c - F_c^5\right)}{{\cal D}_c}\,.\label{mlett:cIII-42}
\eea
\eml

\noindent 
The correspondence between the sets of graphs $a, b, c$, in Fig.10, and the partial amplitudes $T_a, T_b, T_c$ is self explanatory. Furthermore,

\bml
\label{III-43}
\bea
&&F_a \equiv \left[{\bar{u}}(\vec{p}\,'_1,\epsilon'_1) u(\vec{p}_1,\epsilon_1)\right] \left[{\bar{u}}(\vec{p}\,'_2,\epsilon'_2) u(\vec{p}_2,\epsilon_2)
\right]\,,\label{mlett:aIII-43}\\
&&F_a^5 \equiv \left[{\bar{u}}(\vec{p}\,'_1,\epsilon'_1) \gamma^5 u(\vec{p}_1,\epsilon_1)\right] \left[{\bar{u}}(\vec{p}\,'_2,\epsilon'_2) \gamma^5 u(\vec{p}_2,\epsilon_2)\right]\,,\label{mlett:bIII-43}\\
&&{\cal D}_a \equiv \left(p'_1 - p_1\right)^2 \, - \, m^2\,+\,i\epsilon \,,\label{mlett:cIII-43}\\
&&F_b \equiv \left[{\bar{u}}(\vec{p}\,'_1,\epsilon'_1) u(\vec{p}_2,\epsilon_2)\right] \left[{\bar{u}}(\vec{p}\,'_2,\epsilon'_2) u(\vec{p}_1,\epsilon_1)\right]\,,\label{mlett:dIII-43}\\
&&F_b^5 \equiv \left[{\bar{u}}(\vec{p}\,'_1,\epsilon'_1) \gamma^5 u(\vec{p}_2,\epsilon_2)\right] \left[{\bar{u}}(\vec{p}\,'_2,\epsilon'_2) \gamma^5 u(\vec{p}_1,\epsilon_1)\right]\,,\label{mlett:eIII-43}\\
&&{\cal D}_b \equiv \left(p'_1 - p_2\right)^2 \, - \, m^2\,+\,i\epsilon \,, \label{mlett:fIII-43}\\
&&F_c \equiv \left[{\bar{u}}(\vec{p}\,'_1,\epsilon'_1) v(\vec{p}\,'_2,\epsilon'_2)\right] \left[{\bar{v}}(\vec{p}_2,\epsilon_2) u(\vec{p}_1,\epsilon_1)
\right]\,,\label{mlett:gIII-43}\\
&&F_c^5 \equiv \left[{\bar{u}}(\vec{p}\,'_1,\epsilon'_1) \gamma^5 v(\vec{p}\,'_2,\epsilon'_2)\right] \left[{\bar{v}}(\vec{p}_2,\epsilon_2) \gamma^5 u(\vec{p}_1,\epsilon_1)\right]\,,\label{mlett:hIII-43}\\
&&{\cal D}_c \equiv \left(p_1 + p_2\right)^2 \, - \, m^2\,+\,i\epsilon \,, \label{mlett:iIII-43}
\eea
\eml

\be
\label{III-44}
K\,=\, \frac{1}{\pi^2}\,\frac{g^2}{(2\pi)^4}\,\frac{m^2}{\sqrt{\omega(\vec{p}\,'_1)\omega(\vec{p}\,'_2)\omega(\vec{p}_1)\omega(\vec{p}_2)}}\,,
\ee

\noindent 
and $\omega(\vec{p}) \equiv {\sqrt{\vec{p}^{\,2} + m^2}}$. Here, the $u$'s and the $v$'s are, respectively, complete sets of positive and negative energy solutions of the free  Dirac equation. Besides orthogonality and completeness conditions they also obey

\bml
\label{III-45}
\bea
&&C\,{\bar u}^T(\vec{p}, \epsilon)\,=\,v(\vec{p}, \epsilon)\,,\label{mlett:aIII-45}\\
&&C\,{\bar v}^T(\vec{p}, \epsilon)\,=\,u(\vec{p}, \epsilon)\,,\label{mlett:bIII-45}
\eea
\eml

\noindent
where $C \equiv i \gamma^2 \gamma^0$ is the charge conjugation matrix and ${\bar u}^T$ (${\bar v}^T$) denotes the transpose of ${\bar u}$ (${\bar v}$). Explicit expressions for these solutions can be found in Ref.\cite{Muller1}.   

Now, Majorana particles and antiparticles are identical and, unlike
the case for Dirac fermions, all diagrams in Fig.1
contribute to the elastic scattering amplitude of two Majorana
quanta. Then, before going further on, we must verify that the
spin-statistics connection is at work. As expected, $T_a + T_b$
undergoes an overall change of sign when the quantum numbers of the
particles in the outgoing (or in the incoming) channel are exchanged
(see Eqs. (\ref{III-42}) and (\ref{III-43})). As for $T_c$, we notice that

\bml
\label{III-46}
\bea
&&{\bar u}(p, \epsilon) v(p', \epsilon')\,=\,-\,{\bar u}(p', \epsilon') v(p, 
\epsilon)\,, \label{mlett:aIII-46}\\
&&{\bar u}(p, \epsilon) \gamma^5 v(p', \epsilon')\,=\,-\,{\bar u}(p', \epsilon') \gamma^5 v(p, \epsilon)\,, \label{mlett:bIII-46}
\eea
\eml

\noindent
are just direct consequences of Eq.(\ref{III-45}). Thus, $T_c$, alone, also changes sign under the exchange of the outgoing (or incoming) particles and, therefore, $T_a + T_b + T_c$ is antisymmetric.

The main purpose in this section is to disentangle the relevant features
of the low energy regime of the noncommutative Wess-Zumino model model. Since noncommutativity breaks Lorentz invariance, we must carry out this task in an specific frame
of reference that we choose to be the center of mass (CM) frame. Here,
the two body kinematics becomes simpler because one has
that $p_1 = (\omega, \vec{p})$, $p_2 = (\omega, -\vec{p})$, $p'_1 = (\omega,
\vec{p}\,')$, $p'_2 = (\omega, -\vec{p}\,')$, $|\vec{p}\,'| = |\vec{p}\,|$, and
$\omega = \omega(\vec{p})$. This facilitates the calculation of all terms of the form

\be
\label{III-47}
\left[ \frac{m}{\pi \omega(\vec{p})}\right]^2\,\frac{\left(F - F^5\right)}{{\cal D}}, 
\ee

\noindent
in Eqs.(\ref{III-43}). By disregarding all contributions of order $(|\vec{p}\,|/m)^2$ and higher, and after some algebra one arrives at

\bml
\label{III-48}
\bea
 T^L_a\,&=&\,-\frac{1}{(2\pi)^4}\,\left(\frac{g}{\pi \, m}\right)^2\, 
\delta_{\epsilon'_1 \epsilon_1}\, \delta_{\epsilon'_2 \epsilon_2}\,
\left[\frac{1}{2}\cos\left( m \Theta_{0j} k^j\right)\,+\,\frac{1}{2}\cos\left(p^i \Theta_{ij}
k^j\right)\right]\,,\label{mlett:aIII-48}\\
T^L_b\,&=&\,\,\frac{1}{(2\pi)^4}\,\left(\frac{g}{\pi \, m}\right)^2\, 
\delta_{\epsilon'_1 \epsilon_2}\, \delta_{\epsilon'_2 \epsilon_1}\,
\left[\frac{1}{2}\cos\left( m \Theta_{0j} k'\,^j\right)\,+\,\frac{1}{2}\cos\left(p^i \Theta_{ij}
k'\,^j\right)\right]\,,\label{mlett:bIII-48}\\
T^L_c\,&=&\,\,\frac{1}{3(2\pi)^4}\, \left(\frac{g}{\pi \, m}\right)^2\,
\left\{\delta_{\epsilon'_1 \epsilon_1}\, \delta_{\epsilon'_2 \epsilon_2}\,
\cos\left(m \Theta_{0j} p^j\right)\cos\left[m \Theta_{0j} \left(p^j - k^j\right)\right]\right.
\nonumber\\
&&\left. -\,\delta_{\epsilon'_1 \epsilon_2}\, \delta_{\epsilon'_2 \epsilon_1}\,
\cos\left(m \Theta_{0j} p^j\right)\cos\left[m \Theta_{0j} \left(p^j-k'\,^j \right)\right]
\right\}\,,\label{mlett:cIII-48}
\eea
\eml

\noindent
where $k^j \equiv p^j - p'\,^j$ ($k'\,^j \equiv p^j + p'\,^j$) denotes the momentum transferred in the direct (exchange) scattering while 
the superscript $L$ signalizes that the above expressions only hold
true for the low energy regime. It is worth mentioning 
that the dominant contributions to $ T^L_a$ and $T^L_b$ are made by those
diagrams in Fig.10$a$ and Fig.10$b$ not containing
the vertices $i \gamma^5 $, while, on the other hand, the dominant
contribution to $ T^L_c $ comes from the diagram in Fig.10$c$ with
vertices $ i \gamma^5 $. Clearly, $T^L_a + T^L_b +T^L_c $ is antisymmetric under the exchange $\epsilon_1'\leftrightarrow \epsilon'_2$, $\vec{p}\,'\rightarrow -\vec{p}\,'$ ($k^j \leftrightarrow k'\,^j$), as it must be. Also notice that, in the CM frame of reference, only the cosine factors introduced by the space-time noncommutativity are present in $T^L_c$.

We look next for the elastic scattering amplitude involving two
$A$-field quanta. The diagrams contributing to this process, in the
lowest order of perturbation theory, are depicted in Fig.11. The
corresponding (symmetric) amplitude, already written in the CM frame of
reference, can be cast as ${\bar R} = -i(2\pi)^4 \delta^{(4)}(p'_1 + p'_2 - p_1 -
p_2)\,{\bar T}$, where $ {\bar T} = {\bar T}_a + {\bar T}_b + {\bar
T}_c$ and

\bml
\label{III-49}
\bea
{\bar T}_a \,&=&\,\,\frac{g^2}{(2\pi)^4}\,\left(\frac{1}{2\pi 
\omega(\vec{p})}\right)^2\,\left[\frac{1}{2}\cos\left( m \Theta_{0j} k^j\right)\,+\,\frac{1}{2}\cos\left(p^i \Theta_{ij}
k^j\right)\right]\, \frac{1}{\bar {\cal D}}_a \,,\label{mlett:aIII-49}\\
{\bar T}_b \,&=&\,\,\frac{g^2}{(2\pi)^4}\,\left(\frac{1}{2\pi 
\omega(\vec{p})}\right)^2\,\left[\frac{1}{2}\cos\left( m \Theta_{0j} k'\,^j\right)\,+\,\frac{1}{2}\cos\left(p^i \Theta_{ij}
k'\,^j\right)\right]\,\frac{1}{\bar {\cal D}}_b, \label{mlett:bIII-49}\\
{\bar T}_c \,&=&\,\,\frac{g^2}{2 (2\pi)^4}\,\left(\frac{1}{2\pi 
\omega(\vec{p})}\right)^2\,\left\{\cos\left(m \Theta_{0j} p^j\right)\cos\left[m \Theta_{0j} \left(p^j - k^j\right)\right]\right.\nonumber\\
&&\left. +
\cos\left(m \Theta_{0j} p^j\right)\cos\left[m \Theta_{0j} \left (p^j-k'\,^j \right)\right]\right\} \,\frac{1}{\bar {\cal D}}_c\,. \label{mlett:cIII-49}
\eea
\eml            

\noindent
As far as the low energy limit is concerned, the main difference
between the fermionic and bosonic scattering processes rests, roughly speaking,
on the structure of the propagators mediating the interaction. Indeed,
the propagators involved in the fermionic amplitude are those of the
fields $A$ and $B$ given at Eq.(\ref{mlett:aIII-3}), namely,

\[
\Delta_{AA}(p)\,=\,\Delta_{BB}(p)\,=\,i\,{\cal D}^{-1}(p)\,=\,\frac{i}{p^2 - m^2 
+ i\epsilon}\,,
\]

\noindent
which, in all the three cases (a, b, and c), yield a nonvanishing
contribution at low energies (see Eqs.(\ref{mlett:cIII-43}), (\ref{mlett:fIII-43}) and (\ref{mlett:iIII-43})). On the other hand, the propagator involved in the bosonic amplitude is that of the $F$-field given at (\ref{mlett:bIII-3}), i.e.,

\[
\Delta_{FF}\,=\,i\,{\bar {\cal D}}^{-1}(p)\,=\,i\, \frac{p^2}{p^2 - m^2 +i\epsilon}\,,
\]

\noindent
which in turns implies that

\bml
\label{III-50}
\bea
&&{\bar {\cal D}}_a^{-1}\,=\,\frac{2\,\biggl|\frac{\vec{p}}{m}\biggr|^2\,\left(1 - \cos \theta \right)}{ 1 + 2\,\biggl|\frac{\vec{p}}{m}\biggr|^2\,\left(1 - \cos \theta \right)}\,=\,{\cal O}\left( \biggl|\frac{\vec{p}}{m}\biggr|^2\right),\label{mlett:aIII-50}\\
&&{\bar {\cal D}}_b^{-1}\,=\,\frac{2\,\biggl|\frac{\vec{p}}{m}\biggr|^2\,\left(1 + \cos \theta \right)}{ 1 + 2\,\biggl|\frac{\vec{p}}{m}\biggr|^2\,\left(1 + \cos \theta \right)}\,=\,{\cal O}\left( \biggl|\frac{\vec{p}}{m}
\biggr|^2\right)\,,\label{mlett:bIII-50}\\
&&{\bar {\cal D}}_c^{-1}\,=\,\frac{4\, +\, 4\,\biggl|\frac{\vec{p}}{m}\biggr|^2\,}{ 3 + 4\,\biggl|\frac{\vec{p}}{m}\biggr|^2}\,=\,\frac{4}{3 }\,\left[1 + {\cal O}\left( \biggl|\frac{\vec{p}}{m}\biggr|^2\right)\right]\,.\label{mlett:c15} 
\eea
\eml

\noindent
Therefore, at the limit where all the contributions of order $(|\vec{p}\,|/m)^2$ become neglectable, the amplitudes ${\bar T}_a$ and ${\bar T}_b$ vanish whereas ${\bar T}_c$ survives and is found to read

\be
\label{III-51}
{\bar T}^L_c =  \frac{1}{6 (2\pi)^4} \left(\frac{g}{\pi m}\right)^2
\cos\left(m \Theta_{0j} p^j\right)\left\{\cos\left[m \Theta_{0j} \left(p^j - k^j\right)\right]
+ \cos\left[m \Theta_{0j} \left(p^j-k'\,^j \right)\right]\right\}.
\ee

We shall next start thinking of the amplitudes in Eqs.(\ref{III-48}) and
(\ref{III-51}) as of scattering amplitudes deriving from a set of
potentials. These potentials are defined as the Fourier transforms,
with respect to the transferred momentum ($\vec{k}$), of the
respective direct scattering amplitudes. This is due to the fact that
the use, in nonrelativistic quantum mechanics, of antisymmetric wave
functions for fermions and of symmetric wave functions for bosons
automatically takes care of the contributions due to exchange
scattering\cite{Sakurai1}.  Whenever the amplitudes depend only on 
$\vec k$  the corresponding Fourier transforms  will be
local, depending only on a relative coordinate $\vec r$. However, if,
as it happens here, the amplitudes depend  not only on $\vec k$ but also on the initial momentum of the scattered
particle $(\vec p)$, the Fourier transforms will be a function of both $\vec r$
and  $\vec p$.  As the  momentum and position operators do
not commute the construction of potential operators from these
Fourier transforms may be jeopardized by ordering problems. In that
situation, we will proceed as follows: In the Fourier transforms of the
amplitudes we promote the relative coordinate and momentum to
noncommuting canonical conjugated variables and then solve possible
ordering ambiguities by requiring hermiticity of the resulting
expression. A posteriori, we shall verify that this is in fact an effective
potential in the sense that its momentum space matrix elements
correctly reproduce the scattering amplitudes  that we had at the
very  start of this construction.

We are, therefore, led to introduce

\be
\label{III-52}
\delta_{\epsilon'_1 \epsilon_1}\, \delta_{\epsilon'_2 \epsilon_2}\,{\cal M}^F(\vec{k}, \vec{p})\,\equiv\,T_a^L(\vec{k}, \vec{p})\,+\,T_{c,dir}^L(\vec{k}, \vec{p})
\ee

\noindent
and

\be
\label{III-53}
{\cal M}^B(\vec{k},\vec{p})\,\equiv \,{\bar T}_{c,dir}^L(\vec{k}, \vec{p})\,,
\ee

\noindent
in terms of which the desired Fourier transforms ($V^F$ and $V^B$) are given by

\be
\label{III-54}
V^{F,B}({\vec r}, \vec{p})\,=\,(2 \pi)^3\,\int d^3k\,{\cal M}^{F,B}(\vec{k}, \vec{p})\,e^{i \vec{k} \cdot {\vec r}}\,.
\ee

\noindent
In the equations above, the superscripts $F$ and $B$ identify,
respectively, the fermionic and bosonic amplitudes and
Fourier transforms. Also, the subscript $dir$ specifies that only the direct
pieces of the amplitudes $T_c^L$ and ${\bar T}_c^L$ enter in the
calculation of the respective ${\cal M}$. Once $V^{F,B}({\vec r},
\vec{p})$ have been found one has to look for their corresponding quantum operators,
${\hat V}^{F,B}\,({\vec R}, {\vec P})$, by performing the replacements
${\vec r} \rightarrow {\vec R}, {\vec p} \rightarrow {\vec P}$, where
$\vec{R}$ and $\vec{P}$ are the Cartesian position and momentum
operators obeying, by assumption, the canonical commutation relations
$\left[R^j , R^l\right] = \left[P^j , P^l\right]= 0$ and $\left[R^j ,
P^l\right]= i\,\delta^{jl}$. By putting all this together one is led
to the Hermitean forms

\bea
\label{III-55}
{\hat V}^F({\vec R}, {\vec P})\,&=&\,-\,\left(\frac{g}{m}\right)^2\,\int \frac{d^3k}{(2 \pi)^3}
\left( e^{ i k^l R^l } \, e^{i k^l \Theta_{lj} P^j } +  e^{i k^l R^l}\, e^{- i k^l \Theta_{lj} P^j} \right)\nonumber\\
\,&-&\,\frac{2}{3}\left(\frac{g}{m}\right)^2\,\left[\delta^{(3)}\left({\vec R} + m {\vec \Theta}\right)\,+\,\delta^{(3)}\left({\vec R} - m {\vec \Theta}\right)\right]\nonumber\\
&+&\,\frac{1}{3}\left(\frac{g}{m}\right)^2\,\left[\,
\delta^{(3)}\left({\vec R} - m {\vec \Theta}\right)\,e^{- 2 i m {\vec \Theta} \cdot {\vec P}}
\,+\,e^{2 i m {\vec \Theta} \cdot {\vec P}}\,\delta^{(3)}\left({\vec R} - m {\vec \Theta}\right)\right]\,,
\eea

\bea
\label{III-56}
{\hat V}^B({\vec R}, {\vec P})\,&=&\,\frac{1}{6}\left(\frac{g}{m}\right)^2\,
\left[\delta^{(3)}\left({\vec R} + m {\vec \Theta}\right)\,+\,\delta^{(3)}
\left({\vec R} - m {\vec \Theta}\right)\right]\nonumber\\
&+&\,\frac{1}{6}\left(\frac{g}{m}\right)^2\,\left[
\delta^{(3)}\left({\vec R} - m {\vec \Theta}\right)\,e^{- 2i m {\vec \Theta} \cdot {\vec P}}
\,+\,e^{2 i m {\vec \Theta} \cdot {\vec P}}\,\delta^{(3)}\left({\vec R} - m {\vec \Theta}\right)\right]\,,
\eea

\noindent
where ${\vec \Theta} \equiv \{\Theta^{0j}, j=1,2,3\}$. Notice that the
magnetic components of $\Theta_{\mu \nu}$, namely $\Theta_{ij}$, only
contribute to ${\hat V}^F$ and that such contribution is free of
ordering ambiguities, since

\be
\label{III-57}
\left[k^l R^l\,,\,k^m \Theta_{mj} P^j\right]\,=\,i\,k^l \,k^m \,\Theta_{mj}\,\delta^{lj}\,=\,0\,,
\ee

\noindent
in view of the antisymmetry of $\Theta_{mj}$. On the other hand, the
contributions to ${\hat V}^F$ and ${\hat V}^B$ originating in the
electric components of $\Theta_{\mu \nu}$, namely $\Theta_{0j}$, are
afflicted by ordering ambiguities. The relevant point is that there
exist a preferred ordering that makes ${\hat V}^{F}$ and ${\hat
V}^{B}$ both Hermitean, for arbitrary $\Theta_{\mu \nu}$. Equivalent
forms to those presented in Eqs.(\ref{III-55}) and (\ref{III-56}) can be
obtained by using

\be
\label{III-58}
\delta^{(3)} \left({\vec R} - m {\vec \Theta}\right)\,\exp 
\left(- 2im {\vec \Theta} \cdot {\vec P}\right)\,=\,\exp \left(- 2im {\vec 
\Theta} \cdot {\vec P}\right)\,\delta^{(3)} \left({\vec R} + m {\vec 
\Theta}\right)\,.
\ee

We shall shortly verify that the matrix elements of the operators
(\ref{III-55}) and (\ref{III-56}) agree with the original scattering amplitudes.
Before that, however, we want to make some observations about
physical aspects of these operators.

We will consider, separately, the cases of space/space ($\Theta_{0j}
=0$) and space-time ($\Theta_{ij} = 0$) noncommutativity. Hence, we
first set $\Theta_{0j} = 0$ in Eqs.(\ref{III-55}) and (\ref{III-56}). As can
be seen, the potential ${\hat V}^B$, mediating the interaction of two
$A$ quanta, remains as in the commutative case, i.e., proportional to
a delta function of the relative distance between them. The same
conclusion applies, of course, to the elastic scattering of two $B$
quanta. In short, taking the nonrelativistic limit also implies in
wiping out all the modifications induced by the space/space
noncommutativity on the bosonic scattering amplitudes. On the
contrary, Majorana fermions are sensitive to the presence of
space/space noncommutativity. Indeed, from Eq.(\ref{III-55}) follows that
${\hat V}^F$ can be split into planar (${\hat V}^F_P$) and nonplanar
(${\hat V}^F_{NP}$) parts depending on whether or not they depend on
$\Theta_{ij}$, i.e.,

\be
\label{III-59}
{\hat V}^F({\vec R}, {\vec P)}\,=\,{\hat V}^F_P(\vec{R}, {\vec P})\,+
\,{\hat V}^F_{NP}(\vec{R}, {\vec P})\,,
\ee

\noindent
with

\bml
\label{III-60}
\bea
{\hat V}^F_P(\vec{R})\,&=&\,-\,\frac{{2}}{3}\,\left(\frac{g}{m}\right)^2\,\delta^{(3)} (\vec{R})\,,\label{mlett:aIII-60}\\
{\hat V}^F_{NP}(\vec{R}, \vec{P})\,&=&\,-\,\left(\frac{g}{m} \right)^2 \int \, \frac{d^3k}{(2 \pi)^3} \left[ \exp \left(i k^l R^l \right) \, \exp \left(i k^l \Theta_{lj} P^j \right)\right.\nonumber\\
&& \left. +  \exp \left(i k^l R^l \right)\, \exp \left(- i k^l \Theta_{lj} P^j \right) \right]\,.\label{mlett:bIII-60}
\eea
\eml

\noindent
For further use in the Schr\"odinger equation, we shall be needing the position representation of ${\hat{V}^F}\left(\vec{R}, \vec{P}\right)$. From (\ref{mlett:aIII-60}) one easily sees that $<\vec{r}\,|{\hat{V}^F}_P|\vec {r}\,'> = - {2}/3 \,(g/ m)^2\,\delta^{(3)}(\vec{r})\,
\delta^{(3)}(\vec{r} - \vec{r}\,')$. On the other hand, for the computation of $<\vec{r}\,|{\hat {V}^F}_{NP}|\vec {r}\,'>$ it will prove convenient to introduce the realization of $\Theta_{ij}$ in terms of the magnetic field $\vec{B}$, i.e.,

\be
\label{III-61}
\Theta_{ij}\,=\,-\,\epsilon_{ijk}\,B^k\,,
\ee

\noindent
where $\epsilon_{ijk}$ is the fully antisymmetric Levi-Civita tensor ($\epsilon^{123} = +1$). After straightforward calculations one arrives at

\be
\label{III-62}
<\vec{r}\,|{\hat {V}^F}_{NP}|\vec {r}\,'>\,=\,- \frac{2}{(2 \pi)^2}\,\left( \frac{g}{ m}\right)^2\,\frac{1}{B^2}\,\delta^{(1)}({\vec r}_{\parallel})\,\delta^{(1)}({\vec r}_{\parallel} - {\vec r}\,'_{\parallel})\,\cos\left[\frac{\left({\vec r}_{\perp} \times {\vec r}\,'_{\perp}\right) \cdot {\vec B}}{B^2}\right]\,.
\ee

\noindent
Here, ${\vec r}_{\parallel}$ (${\vec r}_{\perp}$) denotes the component of ${\vec r}$ parallel (perpendicular) to ${\vec B}$, i.e., ${\vec r}_{\parallel} = ({\vec r} \cdot {\vec B}){\vec B}/B^2$ (${\vec r}_{\perp} = - ({\vec r} \times {\vec B})\times {\vec B}/B^2$). We remark that  the momentum space matrix element

\bea
\label{III-63}
&&<\vec{p}\,'|{\hat V}^F_{NP}|\vec{p} >\,=\,\int d^3r \int d^3r' <\vec{p}\,'|{\vec r}><{\vec r}\,|{\hat V}^F_{NP}|{\vec r}\,'><{\vec r}\,'|\vec{p} >\nonumber\\
&=&\,-\,\frac{1}{(2\pi)^3}\,\left(\frac{g}{m B}\right)^2\,\int d^2r_{\perp}\,
\exp\left(-i {\vec p}\,'_{\perp} \cdot {\vec r}_{\perp}\right) \nonumber\\
&\times&\left\{\delta^{(2)}\left[{\vec p}_{\perp} - \left({\vec r}_{\perp} \times \frac{{\vec B}}{B^2}\right)\right]\,+\,\delta^{(2)}\left[{\vec p}_{\perp} + \left({\vec r}_{\perp} \times \frac{{\vec B}}{B^2}\right)\right]\right\}\nonumber\\
&=&-\,\frac{1}{4 \pi^3}\,\left(\frac{g}{m}\right)^2\,\cos \left[\left({\vec p}_{\perp} \times {\vec p}\,'_{\perp}\right) \cdot {\vec B}\right]
\eea

\noindent
agrees with the last term in (\ref{mlett:aIII-48}), as it should. We also observe that the interaction only takes place at ${\vec r}_{\perp} = \pm {\vec B}\times {\vec p}_{\perp}$. This implies that ${\vec r}_{\perp}$ must also be orthogonal to ${\vec p}_{\perp}$. Hence, in the case of space/space noncommutativity fermions may be pictured as rods oriented perpendicular to the direction of the incoming momentum. Furthermore, the right hand side of this last equation vanishes if either ${\vec p}_{\perp} \times {\vec p}\,'_{\perp} = 0$, or $\left({\vec p}_{\perp} \times {\vec p}\,'_{\perp}\right) \cdot {\vec B} = 0$, or $\vec{p} = \vec{p}_{\parallel}$, or $\vec{p}\,' = \vec{p}\,'_{\parallel}$.

In the Born approximation, the fermion-fermion elastic scattering amplitude ($f^F({\vec p}\,',{\vec p})$) can be computed at once, since $f^F({\vec p}\,',{\vec p}) = -4 \pi^2 m <\vec{p}\,'|{\hat V}^F|\vec{p}>$. In turns, the corresponding outgoing scattering state ($\Phi^{F(+)}_{\vec{p}}(\vec r)$) is found to behave asymptotically ($r \rightarrow \infty$) as follows

\bea
\label{III-64}
&& e^{-iEt}\, \Phi^{F(+)}_{\vec{p}}(\vec r)\,\sim\,\left(\frac{1}{2\pi}\right)^{\frac{3}{2}}\left[ e^{-i \left(E t -\vec{p} \cdot {\vec r}\right)}\,+\,\frac{e^{-i\left(E t -pr\right)}}{r}\,f^F({\vec p}\,',{\vec p})\right]\nonumber\\
& \sim & \left(\frac{1}{2\pi}\right)^{\frac{3}{2}}\left\{e^{-i \left(E t - \vec{p} \cdot {\vec r}\,\right)}+\,\frac{g^2}{3\pi m} \frac{e^{- i(Et - pr)}}{r} \right.\nonumber\\
&&\left. + \frac{g^2}{2\pi m}\,\left [\frac{e^{- i\left[Et - \left(\vec{p}_{\perp} \times \vec{p}\,'_{\perp}\right) \cdot {\vec B} - pr\right]}}{r}\,+\,\frac{e^{- i\left[Et + \left(\vec{p}_{\perp} \times \vec{p}\,'_{\perp}\right) \cdot {\vec B} - pr\right]}}{r}\right]\right\}\,,
\eea

\noindent
where $E = \vec{p}^2/2m$ is the energy of the incoming particle. The right hand side of Eq.(\ref{III-64}) contains three scattered waves. The one induced by the planar part of the potential (${\hat V}^F_P$) presents no time delay. The other two originate in the nonplanar part of the potential (${\hat V}^F_{NP}$) and exhibit time delays of opposite signs and proportional to $\left(\vec{p}_{\perp} \times \vec{p}\,'_{\perp}\right) \cdot {\vec B}$. For instance, for ${\vec B}$ and ${\vec p}$ along the positive Cartesian semiaxis $x^1$ and $x^3$, respectively, one has that   
$\left(\vec{p}_{\perp} \times \vec{p}\,'_{\perp}\right) \cdot {\vec B}\,=\,- 2mEB \sin\theta \sin\phi$, were, $\theta$ and $\phi$ are the scattering and azimuthal angles, respectively. The $\phi$-dependence reflects the breaking of rotational invariance. 

We set next $\Theta_{ij} = 0$, in Eqs(\ref{III-55}) and (\ref{III-56}), and turn into analyzing the case of space-time noncommutativity. The effective potentials are now 

\bea
\label{III-65}
{\hat {\tilde V}}^F({\vec R}, {\vec P})\,&=&\,-\,\frac{2}{3}\left(\frac{g}{m}\right)^2\,\left[\delta^{(3)}\left({\vec R} + m {\vec \Theta}\right)\,+\,\delta^{(3)}\left({\vec R} - m {\vec \Theta}\right)\right]\nonumber\\
&+&\,\frac{1}{3}\left(\frac{g}{m}\right)^2\,\left[\,
\delta^{(3)}\left({\vec R} - m {\vec \Theta}\right)\,e^{- 2 i m {\vec \Theta} \cdot {\vec P}}
\,+\,e^{2 i m {\vec \Theta} \cdot {\vec P}}\,\delta^{(3)}\left({\vec R} - m {\vec \Theta}\right)\right]\,,
\eea

\bea
\label{III-66}
{\hat {\tilde V}}^B({\vec R}, {\vec P})\,&=&\,\frac{1}{6}\left(\frac{g}{m}\right)^2\,\left[\delta^{(3)}\left({\vec R} + m {\vec \Theta}\right)\,+\,\delta^{(3)}\left({\vec R} - m {\vec \Theta}\right)\right]\nonumber\\
&+&\,\frac{1}{6}\left(\frac{g}{m}\right)^2\,\left[
\delta^{(3)}\left({\vec R} - m {\vec \Theta}\right)\,e^{- i 2m {\vec \Theta} \cdot {\vec P}}
\,+\,e^{ i 2m {\vec \Theta} \cdot {\vec P}}\,\delta^{(3)}\left({\vec R} - m {\vec \Theta}\right)\right]\,,
\eea

\noindent
where the slight change in notation (${\hat V} \rightarrow {\hat {\tilde V}}$) is for avoiding confusion with the previous case. As before, we look first for the fermionic and bosonic elastic scattering amplitudes and then construct the asymptotic expressions for the corresponding scattering states. Analogously to (\ref{III-63}) and (\ref{III-64}) we find that

\bea
\label{III-67}
<\vec{p}\,'|{\hat {\tilde V}}^F_{NP}|\vec{p} > &=& 
-\frac{1}{12  \pi^3}\,\left( \frac{g}{m}\right )^2\left\{2 \cos \left[m {\vec \Theta} \cdot \left( \vec{p} - \vec{p}\,'\right)\right]\,-\,\cos \left[m {\vec \Theta} \cdot \left( \vec{p} + \vec{p}\,'\right)\right]\right\}\,
\eea

\noindent
and

\bea
\label{III-68}
e^{-iEt}\,{\tilde \Phi}_{\vec{p}}^{F (+)}{(\vec r)}\,&=&\,\left(\frac{1}{2\pi}\right)^{\frac{3}{2}}
\left \{e^{-i\left(Et - \vec{p} \cdot {\vec r}\right)}\right.\nonumber\\
&& \left.+\,\frac{g^2}{3 m \pi r}\,\left [{e^{-i\left[Et - m {\vec \Theta} \cdot \left( \vec{p} - \vec{p}\,'\right) - pr \right]}}\,+\,  
{e^{-i\left[Et + m {\vec \Theta} \cdot \left( \vec{p} - \vec{p}\,'\right) - pr \right]}}\right ]\right. \nonumber\\
&& \left. -\,\frac{g^2}{6 m \pi r}\,\left [{e^{-i\left[Et - m {\vec \Theta} \cdot \left( \vec{p} + \vec{p}\,'\right) - pr \right]}}\,+\,  
{e^{-i\left[Et + m {\vec \Theta} \cdot \left( \vec{p} + \vec{p}\,'\right) - pr \right]}}\right ]\right\}\,.
\eea

\noindent
in accordance with the low energy limit of the relativistic calculations. As for the bosons, the potential in Eq.(\ref{III-66}) leads to

\bea
\label{III-69}
<\vec{p}\,'|{\hat {\tilde V}}^B_{NP}|\vec{p} >\, &=&\frac{1}{24  \pi^3}\,\left( \frac{g}{m}\right )^2\,\,\left\{ \cos \left[m {\vec \Theta} \cdot \left( \vec{p} - \vec{p}\,'\right)\right]\,+\,\cos \left[m {\vec \Theta} \cdot \left( \vec{p} + \vec{p}\,'\right)\right]\right\}\,
\eea

\noindent
and

\bea
\label{III-70}
e^{-iEt}\,{\tilde \Phi}_{\vec{p}}^{B (+)}{(\vec r)}\,&=&\,\left(\frac{1}{2\pi}\right)^{\frac{3}{2}}
\Bigl \{e^{-i\left(Et - \vec{p} \cdot {\vec r}\right)}\Bigr.\nonumber\\
&& \left.-\,\frac{g^2}{12 m \pi r}\,\left [{e^{-i\left[Et - m {\vec \Theta} \cdot \left( \vec{p} - \vec{p}\,'\right) - pr \right]}}\,+\,  
{e^{-i\left[Et + m {\vec \Theta} \cdot \left( \vec{p} - \vec{p}\,'\right) - pr \right]}}
\right. \right. \nonumber\\[12pt]
&& \left. \left. +\,{e^{-i\left[Et - m {\vec \Theta} \cdot \left( \vec{p} + \vec{p}\,'\right) - pr \right]}}\,+\,  
{e^{-i\left[Et + m {\vec \Theta} \cdot \left( \vec{p} + \vec{p}\,'\right) - pr \right]}}\right ]\right \}\,.
\eea

We stress that, presently, the interaction only takes place at ${\vec
r} = \pm (\vec{p} - \vec{p}\,')/m^2$ and ${\vec r} = \pm (\vec{p} + \vec{p}\,')/m^2$
(see Eqs.(\ref{III-67}) and (\ref{III-69})). As consequence, particles in
the forward and backward directions behave as rigid rods oriented
along the direction of the incoming momentum $\vec{p}$.  Furthermore, each
scattering state (see Eqs.(\ref{III-68}) and (\ref{III-70})) describes four
scattered waves. Two of these waves are advanced, in the sense that
the corresponding time delay is negative, analogously to what was
found in \cite{Seiberg2}.

\newpage

\section{Acknowledgments}

This work was partially supported by Conselho Nacional de
Desenvolvimento Cient\'\i fico e Tecnol\'ogico (CNPq) and by PRONEX under contract CNPq 66.2002/1998-99.

\newpage

\input{epsf.tex}
\begin{figure}[h]
\centerline{\epsfbox{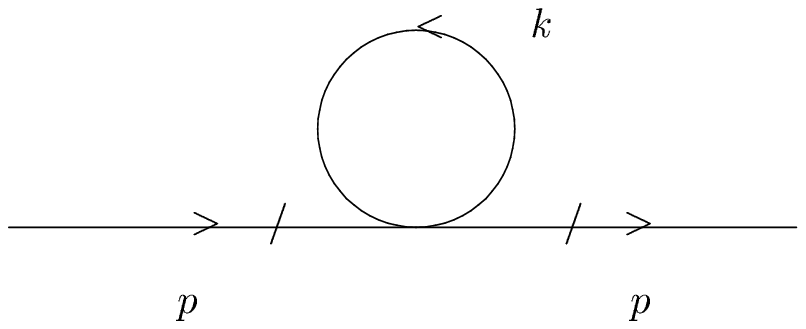}}
\caption{{\it One--loop contributions to $\tilde{\Gamma}^{(2)}(p)$. The slash denotes amputed lines}}
\label{Fig.1}
\end{figure}
\begin{figure}[h]
\centerline{\epsfbox{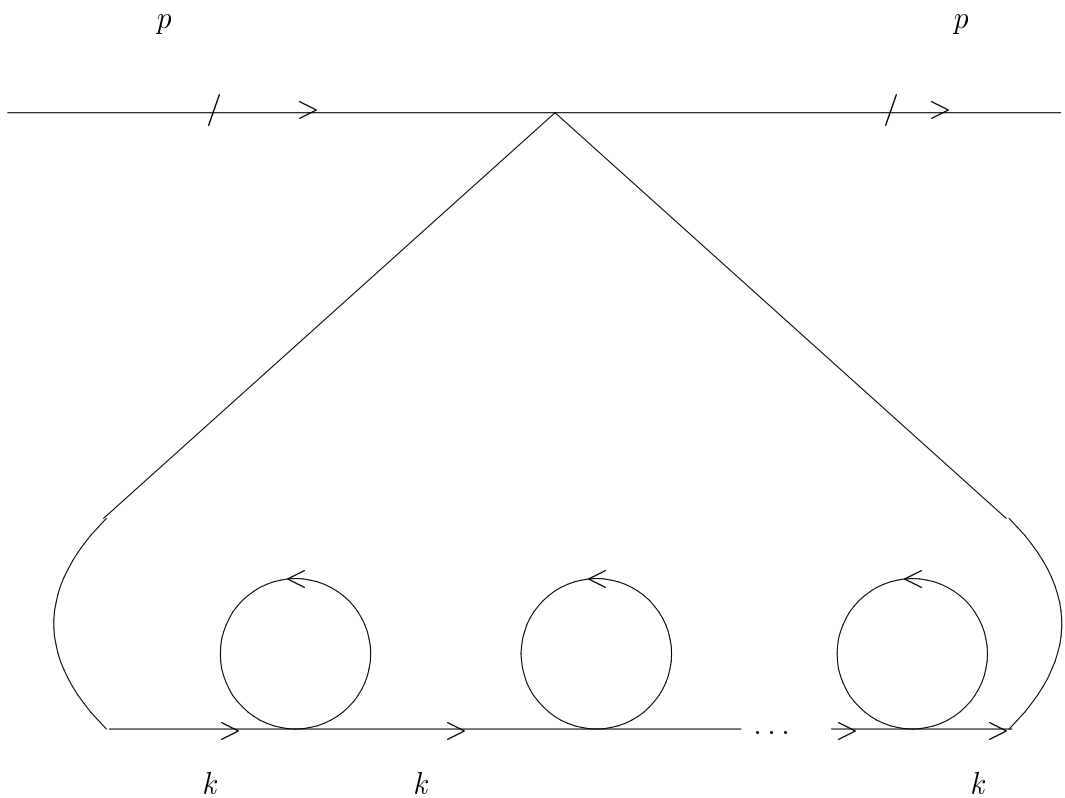}}
\caption{{\it Insertion of $\Sigma(p)$ in higher order diagrams.}}
\label{Fig.2}
\end{figure}
\begin{figure}[h]
\centerline{\epsfbox{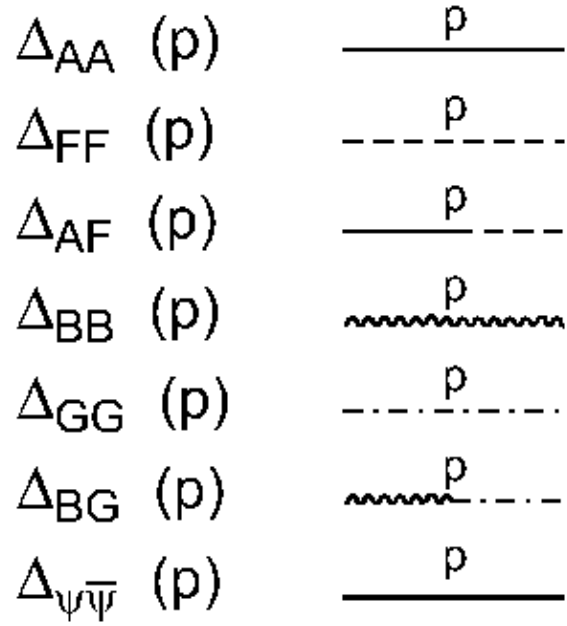}}
\caption{{\it Graphical representation for the propagators.}}
\label{Fig.3}
\end{figure}
\begin{figure}[h]
\centerline{\epsfbox{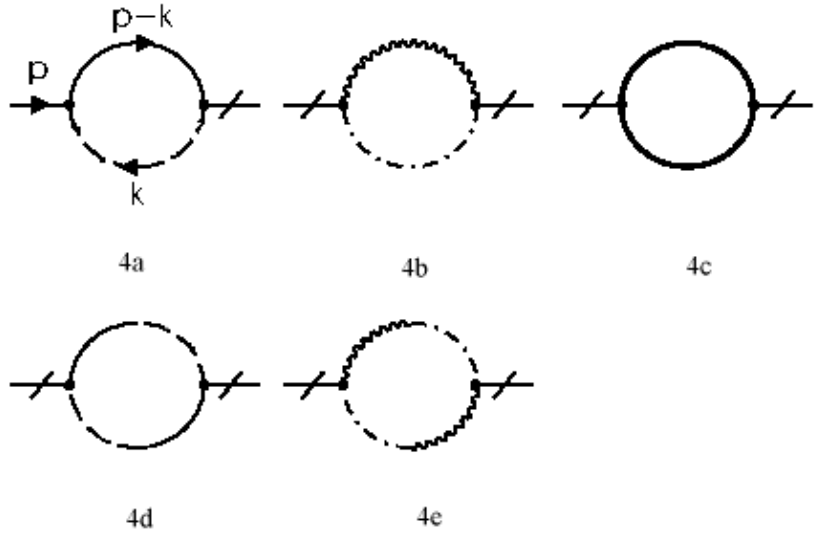}}
\caption{{\it One--loop contributions to the self-energy of the $A$ field.}}
\label{Fig.4}
\end{figure}
\begin{figure}[h]
\centerline{\epsfbox{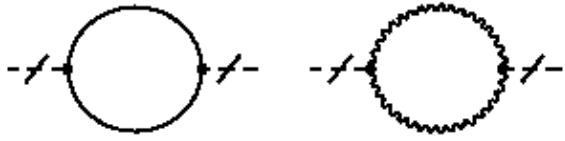}}
\caption{{\it One--loop corrections to the two point function of the auxiliary
field $F$. }}
\label{Fig.5}
\end{figure}
\begin{figure}[h]
\centerline{\epsfbox{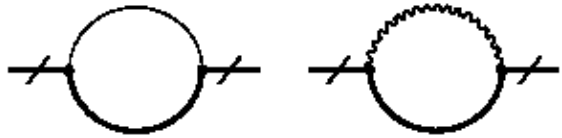}}
\caption{{\it One--loop contributions to the self-energy of the spinor field
$\psi$.}}
\label{Fig.6}
\end{figure}
\begin{figure}[h]
\centerline{\epsfbox{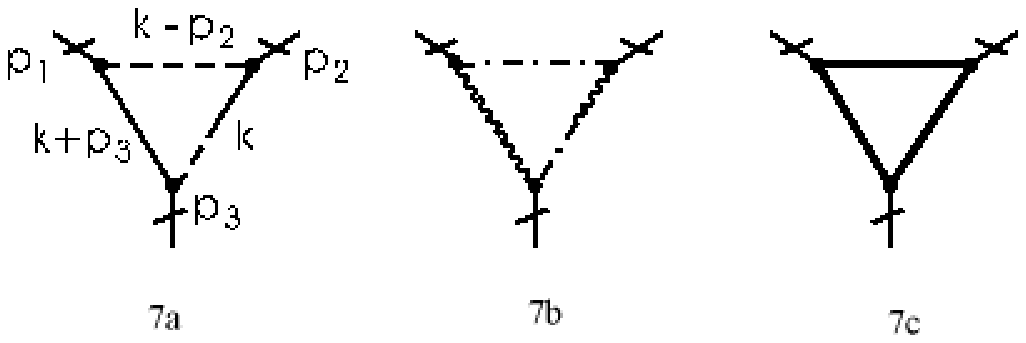}}
\caption{{\it Divergent graphs contributing to the three point function  of 
the $A$ field.}}
\label{Fig.7}
\end{figure}
\begin{figure}[h]
\centerline{\epsfbox{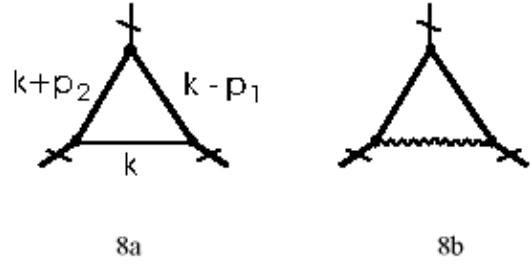}}
\caption{{\it One--loop contributions to the three point function 
$\Gamma(A\overline \psi\psi)$}}
\label{Fig.8}
\end{figure}
\begin{figure}[h]
\centerline{\epsfbox{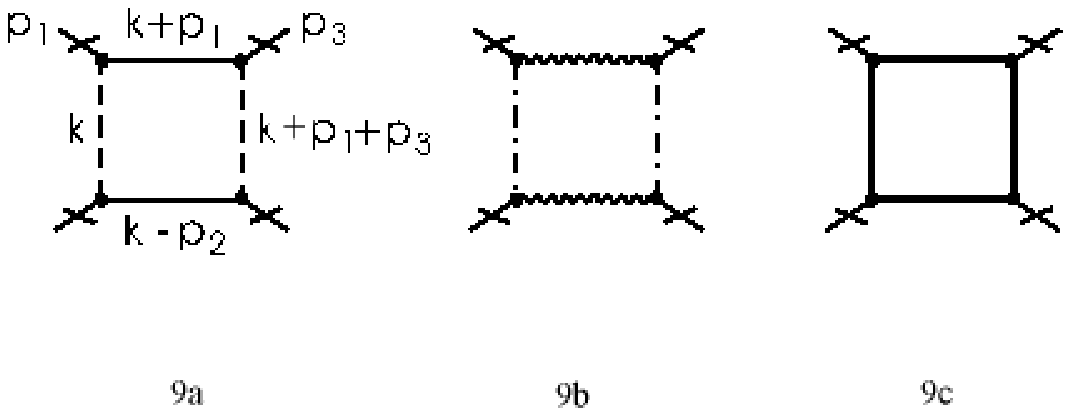}}
\caption{{\it Divergent graphs contributing to the four point function of the
$A$ field.}}
\label{Fig.9}
\end{figure}
\begin{figure}[h]
\centerline{\epsfbox{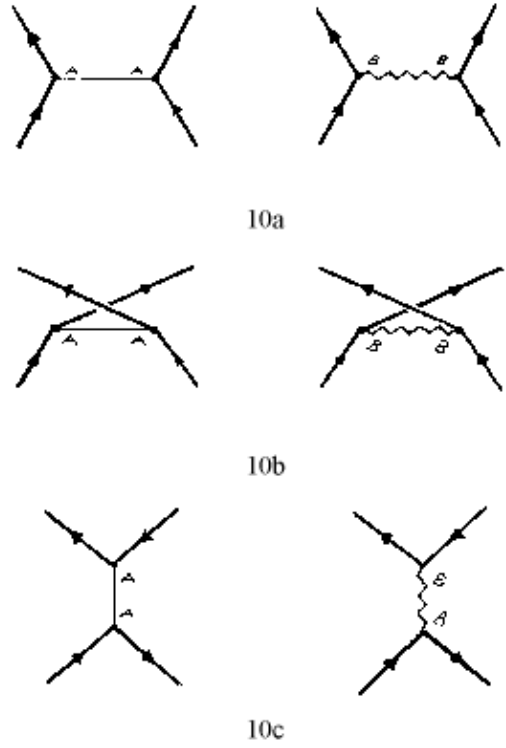}}
\caption{{\it Lowest order graphs contributing to the scattering of two Majorana spinors.}}
\label{Fig.10}
\end{figure}
\begin{figure}[h]
\centerline{\epsfbox{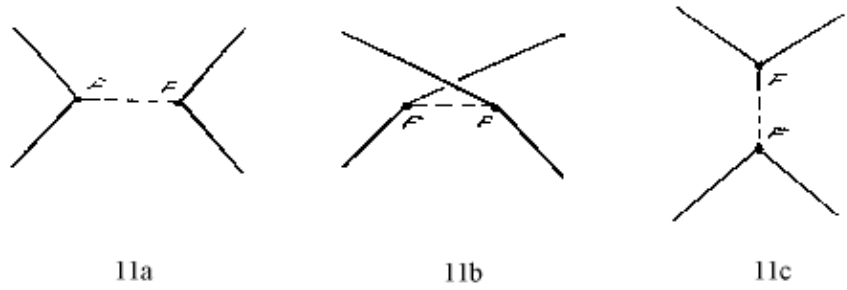}}
\caption{{\it Lowest order graphs contributing to the scattering of two $A$-quanta.}}
\label{Fig.11}
\end{figure}

\end{document}